\pdfoutput=1 
\documentclass[a4paper,11pt]{article}
\PassOptionsToPackage{table}{xcolor}

\usepackage{jheppub} 

\usepackage[T1]{fontenc} 
\usepackage{tangocolors}
\usepackage{tikz}
\usetikzlibrary{arrows,snakes,backgrounds}
\usepackage{amssymb}
\usetikzlibrary{decorations.pathmorphing,backgrounds,shapes,arrows,shadows}
\tikzset{zigzag/.style={decorate,decoration=zigzag}}
\tikzset{snake it/.style={decorate, decoration=snake}}
\makeatletter
\def\@hex@@Hex#1%
 {\if a#1A\else \if b#1B\else \if c#1C\else \if d#1D\else
  \if e#1E\else \if f#1F\else #1\fi\fi\fi\fi\fi\fi \@hex@Hex}
\makeatother

\usepackage[all]{xy}
\usepackage[percent]{overpic}
\usepackage{slashed}
\usepackage{wrapfig}
\usepackage{tabu}
\usepackage{diagbox}
\usepackage{mathrsfs,amsmath,amssymb,amsthm,amsfonts,tikz,graphicx,accents,hyperref, color}
\usepackage{dsfont,epiolmec, latexsym, stmaryrd, comment}
\usepackage{slashed,ccaption}
\usepackage{mathrsfs, calligra}
\usepackage{leftidx}
\usepackage{import}
\usepackage{multirow}
\usepackage{amsfonts}
\usepackage{pifont}
\usepackage{tabularx}
\usepackage{cancel}
\usepackage[utf8]{inputenc}
\usetikzlibrary{intersections,calc}
\usepackage{ifthen}
\usepackage{amsmath}
\usepackage{cancel}
\usepackage{caption}

\usepackage{array}
\hypersetup{ linktoc=all,
    colorlinks, linkcolor={palatinateblue},
    citecolor={brightpink}, urlcolor={amaranth}
}

\graphicspath{{Images/}}

\renewcommand{\d}[1]{\ensuremath{\operatorname{d}\!{#1}}}

\def\sideremark#1{\ifvmode\leavevmode\fi\vadjust{\vbox to0pt{\vss
 \hbox to 0pt{\hskip\hsize\hskip1em
 \vbox{\hsize2cm\tiny\raggedright\pretolerance10000
 \noindent #1\hfill}\hss}\vbox to8pt{\vfil}\vss}}}%
\DeclareSymbolFont{extraup}{U}{zavm}{m}{n}
\DeclareMathSymbol{\varheart}{\mathalpha}{extraup}{86}
\DeclareMathSymbol{\vardiamond}{\mathalpha}{extraup}{87}
\makeatletter
\renewcommand*{\@fnsymbol}[1]{\ensuremath{\ifcase#1\or \clubsuit \or \vardiamond \or \varheart\or
    \spadesuit\or \mathparagraph\or \|\or **\or \dagger\dagger
    \or \ddagger\ddagger \else\@ctrerr\fi}}
\makeatother

\definecolor{rosy}{RGB}{230,235,252}
\definecolor{myframetitle}{RGB}{90,89,170}
\definecolor{myblocktitle}{RGB}{140,185,249}
\definecolor{mytitle}{RGB}{10,80,26}

\definecolor{darkgreen}{RGB}{27,130,45}
\definecolor{darkblue}{rgb}{0,0,0.3}
\definecolor{darkred}{rgb}{0.7,0,0}

\definecolor{light gray}{RGB}{220,220,220}
\definecolor{dark purple}{RGB}{108,0,217}
\definecolor{pink}{RGB}{190,20,100}
\definecolor{orang}{RGB}{193,63,0}
\definecolor{green}{RGB}{11,98,17}
\definecolor{darkpink}{RGB}{153,0,76}
\definecolor{bluegreen}{RGB}{0,102,102}
\definecolor{greenlagan}{RGB}{0,102,0}
\definecolor{redgreen}{RGB}{102,102,0}
\definecolor{Redgreen}{RGB}{153,76,0}
\definecolor{vividviolet}{rgb}{0.62, 0.0, 1.0}
\definecolor{amaranth}{rgb}{0.9, 0.17, 0.31}
\definecolor{palatinateblue}{rgb}{0.15, 0.23, 0.89}
\definecolor{brightpink}{rgb}{1.0, 0.0, 0.5}
\definecolor{cornflowerblue}{rgb}{0.39, 0.58, 0.93}
\definecolor{deepcarminepink}{rgb}{0.94, 0.19, 0.22}
\definecolor{radicalred}{rgb}{1.0, 0.21, 0.37}

\newcommand\pnote[1]{\textcolor{magenta}{\bf [P:\,#1]}}
\newcommand\hnote[1]{\textcolor{blue}{\bf [HA:\,#1]}}

\newcommand\vnote[1]{\textcolor{cyan}{\bf [V:\,#1]}}
\newcommand\HYnote[1]{\textcolor{green}{\bf [HY:\,#1]}}

\newcommand\bc[1]{\boldsymbol{\mathcal{#1}}}

\newcommand{\cc}{{\mathcal C }}

\DeclareFontFamily{OT1}{rsfs}{}

\DeclareFontShape{OT1}{rsfs}{m}{n}{ <-7> rsfs5 <7-10> rsfs7 <10->rsfs10}{} 

\DeclareMathAlphabet{\mycal}{OT1}{rsfs}{m}{n}

\newcommand{\be}{\begin{equation}}
\newcommand{\ee}{\end{equation}}
\newcommand{\bea}{\begin{eqnarray}}
\newcommand{\eea}{\end{eqnarray}}
\makeatletter \@addtoreset{equation}{section}

\begin{document}


\newcommand{\mytitle}{\centerline{\LARGE{Symmetries at Causal Boundaries in 2D and 3D Gravity}}}\vskip 3mm 

\title{{\mytitle}}

\author[a,b]{H.~Adami}
\author[c]{, Pujian~Mao}
\author[d]{, M.M.~Sheikh-Jabbari}
\author[d,e]{, V.~Taghiloo}
\author[b]{, H.~Yavartanoo}
\affiliation{$^a$ Yau Mathematical Sciences Center, Tsinghua University, Beijing 100084, China}
\affiliation{$^b$ Beijing Institute of Mathematical Sciences and Applications (BIMSA), Huairou District, Beijing 101408, P. R. China}

\affiliation{$^c$ Center for Joint Quantum Studies and Department of Physics,
School of Science, Tianjin University, 135 Yaguan Road, Tianjin 300350, China
}

\affiliation{$^d$ School of Physics, Institute for Research in Fundamental
Sciences (IPM), P.O.Box 19395-5531, Tehran, Iran}
\affiliation{$^e$ Department of Physics, Institute for Advanced Studies in Basic Sciences (IASBS),
P.O. Box 45137-66731, Zanjan, Iran}

\emailAdd{hamed.adami@bimsa.cn, 
pjmao@tju.edu.cn, 
jabbari@theory.ipm.ac.ir,  v.taghiloo@iasbs.ac.ir, yavar@bimsa.cn
}

\abstract{We study  $2d$ and $3d$  gravity theories on spacetimes with causal (timelike or null) codimension one boundaries while allowing for variations in the position of the boundary. We construct the corresponding solution phase space and specify boundary degrees freedom by analysing boundary (surface) charges labelling them. We discuss $Y$ and $W$ freedoms and change of slicing in the solution space. For $D$ dimensional case we find $D+1$  surface charges, which are generic functions over the causal boundary. We show that there exist solution space slicings in which the charges are integrable. For the $3d$ case there exists an integrable slicing where charge algebra takes the form of Heisenberg $\oplus\ {\cal A}_3$ where ${\cal A}_3$ is two copies of Virasoro at Brown-Henneaux central charge for AdS$_3$ gravity and BMS$_3$ for the $3d$ flat space gravity. 
}
\maketitle

\section{Introduction}

In many physically relevant situations one needs to formulate physics on spacetime with boundaries. A boundary is a codimension one surface which cuts spacetime into two parts or it  marks the asymptotic regions where spacetime resides only in one side of this codimension one surface. A boundary can be a physical object or be a hypothetical surface in the spacetime with prescribed features. In general it can be spacelike (like Cauchy surfaces),  null (like asymptotic region of a flat spacetime or horizon of a stationary black hole) or timelike (like AdS causal boundary or a codimension one brane in the spacetime or walls of a cubic box in the spacetime). Depending on the physical problem and the properties of the boundary we deal with different situations which may be formulated quite differently.  

In presence of a boundary we typically need to add boundary degrees of freedom (BDOF), degrees of freedom which reside on the boundary and do not propagate into the bulk. For the case of a spacelike boundary, the boundary data is fully encoded in the Cauchy (initial) data. In the case of a causal boundary, which will be the focus of our study here, the BDOF can be dynamical ones, have their own independent dynamics while interact with the bulk degrees of freedom. The role of BDOF is in part to ensure a prescribed boundary condition on the bulk fields. This, however, is not generically enough to completely fix the BDOF and their dynamics.  The first step in formulating the boundary dynamics is to identify the BDOF. In gauge or gravity theories this identification for hypothetical boundaries can be done through gauge or diffeomorphisms which nontrivially act on the boundary. 

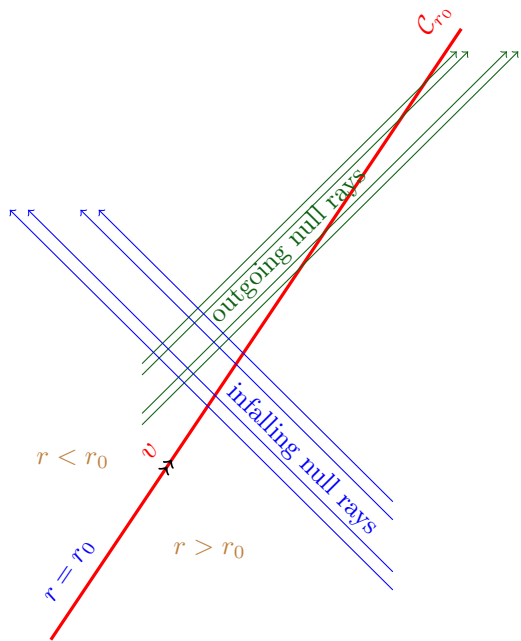
\begin{figure}
\def \L {3.0}
    \centering
\begin{tikzpicture}
  \draw[very thick,red] (-\L,-\L) coordinate (b) -- (0.8*\L,1.7*\L) coordinate (t);
\draw[green,->] (-0.6*\L,-0.05*\L) -- (1.05*\L,1.6*\L);  
\draw[green,->] (-0.6*\L,0*\L) -- (1*\L,1.6*\L);   
\draw[green,->] (-0.6*\L,0.17*\L) -- (0.83*\L,1.6*\L);   
\draw[green,->] (-0.6*\L,0.22*\L) -- (0.78*\L,1.6*\L);  
\draw[green] (0.4*\L,1.1*\L) node[left, rotate=+45] (scrip) {\small{outgoing null rays}};
\draw[blue,->] (0.5*\L,-0.39*\L) -- (-0.79*\L,0.9*\L);           
  \draw[blue,->] (0.5*\L,-0.47*\L) -- (-0.87*\L,0.9*\L);
  \draw[blue,->] (0.5*\L,-0.7*\L) -- (-1.1*\L,0.9*\L);
  \draw[blue,->] (0.5*\L,-0.78*\L) -- (-1.18*\L,0.9*\L);
   \draw[black,thick,->] (-0.5*\L,-0.25*\L)--(-0.49*\L,-0.235*\L); \draw[black,thick,->] (-0.48*\L,-0.22*\L)--(-0.47*\L,-0.205*\L);
   \draw[red] (-0.52*\L,-0.1*\L) node[left, rotate=56] (scrip) {{$v$}};
 \draw[blue] (0.45*\L,-0.55*\L) node[left, rotate=-45] (scrip) {\small{infalling null rays}};
    \draw[blue] (-0.8*\L,-0.53*\L) node[left, rotate=56] (scrip) {\small{$r=r_0$}}; 
    \draw[red] (0.75*\L,1.85*\L) node[left, rotate=56] (scrip) {\small{${\cal C}_{r_0}$}};
    \draw[brown] (-0.1*\L,-0.6*\L) node[left] (scrip) {\small{$r>r_0$}};
    \draw[brown] (-0.7*\L,-0.2*\L) node[left] (scrip) {\small{$r<r_0$}};
\end{tikzpicture}
     \caption{Depiction of a causal boundary at an arbitrary $r=r_0$. We want to formulate physics in the `outside' $r\geq r_0$ region and excise the $r<r_0$ part. Unlike the null boundary case of \cite{Adami:2021nnf}, here we can have both infalling and outgoing null rays passing through the causal boundary. In our setting we allow for `local boosts' which encode fluctuations of the causal boundary ${\cal C}_{r_0}$, this adds one more surface charge compared to the null boundary case. For the $2d$ and $3d$ cases we consider here, however, we do not have bulk modes and hence there are no infalling or outgoing null rays.}
    \label{fig:causal-boundary}
\end{figure}

Here, we will be interested in 2 and 3 dimensional gravity theories and causal (timelike or null) boundaries, and consider cases where the boundary is a hypothetical surface which cuts the spacetime into two parts; we will not consider the asymptotic boundaries, see Fig. \ref{fig:causal-boundary}. 
Mainly motivated by questions regarding black holes, in a series of previous papers we have analyzed a similar problem for null boundaries which model horizons: In \cite{Adami:2020ugu} we considered 2  dimensional Einstein  dilaton gravity and 3 dimensional Einstein-$\Lambda$ theory, in \cite{Adami:2021sko} studied 3 dimensional topologically massive gravity and in \cite{Adami:2021nnf} we considered generic $D$ dimensional pure Einstein gravity. In these settings we do not impose  specific boundary conditions on the null surface and allow for all possible fluctuations which leave the boundary a null surface. Through a thorough boundary symmetry analysis within covariant phase space formalism, we established that $D$ surface charges as functions over the $D-1$ dimensional null boundary. These charges label the BODF, which are one $D-1$ vector and one scalar field,  on the null boundary. These charges satisfy an algebra which depends on the slicing used for the solution phase space, and the BDOF fall into representations of this algebra. 

For the generic causal boundary where the boundary is allowed to fluctuate along the  transverse directions, here we establish that for $D=2,3$ dimensions,  there are $D+1$ charges, associated with two scalar fields and a vector field along the $D-1$ dimensional boundary. The extra scalar compared to the null boundary case is associated with the fluctuations transverse to the boundary; in the Fig. \ref{fig:causal-boundary} this corresponds to `transverse supertranslations' which encode fluctuations of the causal codimension 1 boundary ${\cal C}_{r_0}$. We work out boundary symmetry algebras and charges and the associated central terms. While the algebra of charges generically depend on the slicing of the phase space, we show that there exists a slicing in which the two scalar charges satisfy a Heisenberg algebra. 

At a technical level to perform the analysis we need to address various issues within the covariant phase space formalism, most notably the so-called $W$ and $Y$ ambiguities, which as we discuss, we prefer to call them ``freedom'' in the definition of charges instead of ambiguities. In particular, we fix the $Y$-freedom to fulfill the physical expectation that the symplectic form, surface charges and their algebra are independent of the position of the causal boundary.\footnote{In \cite{Ruzziconi:2020wrb,Geiller:2021vpg} a similar problem for the $2d$ dilaton gravity and the AdS$_3$ gravity with asymptotic boundary is studied. It was shown in \cite{Geiller:2021vpg} that one can get 4 codimension 1 charges and the charges can be made finite (and essentially independent of the AdS$_3$ radial coordinate) upon an appropriate choice for $Y$-freedom.}   The covariant phase space method yields charge variations over the solution space. We use Barnich-Troessaert  (BT) method \cite{Barnich:2011mi} to separate the charge variation into an integrable part and a non-integrable ``flux'' part. We fix the $W$-freedom such that the integrable charges obtained from the BT method to be identical to the Noether charge. We argue that this should always be possible and is not limited to the specific examples we analyse. This is compatible with and confirms the proposal made in \cite{Freidel:2021cjp}. Moreover, as conjectured in \cite{Adami:2020ugu} and established in \cite{Adami:2021nnf}, integrability of charges in general depends on the phase space slicing. In particular, for the $D=2,3$ cases where there is no propagating bulk degree of freedom, we expect there should be slicings of the phase space in which charges are integrable. We present  integrable slicings for the $D=2,3$ examples and discuss the charge algebra in these slicings.

This paper is organized as follows. In section \ref{sec:2}, we analyze solution space and boundary symmetries, charges and algebra for a $2d$ scalar-tensor gravity theory in presence of a causal (null or timelike) boundary. In section \ref{sec:3''}, we  study a similar problem for the $3d$ Einstein-$\Lambda$ theory. In section \ref{sec:discussion}, we discuss further  our results and give an outlook. In  appendix \ref{sec:bdryCharge}, we give a quick review of covariant phase space formalism and computation of symplectic form, charge variations, the Barnich-Troessaert method, computation of Noether charges and the $W$ and $Y$ freedoms. {In  appendix \ref{sec:change of slicing}, we review two other technical points we have used in our analysis, the notion of adjusted bracket and the change of slicing over the solution space. }
\section{Causal Boundary Symmetries, $2d$ Case}\label{sec:2}

The two dimensional dilaton gravity we consider is described by the action\footnote{See \cite{Grumiller:2021cwg} for more general class of $2d$ gravity theories and surface charge analysis.}  
\begin{equation}\label{2D-action}
    S= \frac{1}{16\pi G} \int \d{}^2 x \sqrt{-g}  \left[ \Phi \, R - X(\Phi) \right]
\end{equation}
where ${R}$ is Ricci scalar, $\Phi$ is the scalar field and $X(\Phi)$ is the potential term. 
The field equations for this action are given by
\begin{subequations}\label{EOM}
\begin{align}
    &0=\nabla_{\mu}\nabla_{\nu}\Phi-\frac{1}{2}g_{\mu\nu}\Box \Phi ,\label{eom-1}\\
    & 0=\Box \Phi+ X,\label{eom-2}\\
    &0=R-\frac{\d X}{\d \Phi}. \label{eom-3}
\end{align}
\end{subequations}
The Jackiw-Teitelboim (JT) gravity is within this family with ${X(\Phi)=2\Lambda\Phi}$ {, where $\Lambda$ is a constant}. Another member of this $2d$ gravity family is the one obtained from dimensional reduction of ${D}$ dimensional Einstein-$\Lambda$ gravity over an ${S^{D-2}}$. Explicitly,  consider the reduction anstaz, 
\begin{equation}\label{line-element-HDS}
    \d {} \tilde{s}^2= \Phi^{-\frac{D-3}{(D-2)}}g_{\mu\nu}\d x^{\mu}\d x^\nu + \Phi^{\frac{2}{(D-2)}}\, \d{}\Omega_{D-2}^2,
\end{equation}
where $\mu,\nu=0,1$ and $\d{}\Omega_{D-2}^2$ is the metric of a round unit radius $S^{D-2}$. Upon reduction one obtains a 2 dimensional action \eqref{2D-action} with $X(\Phi)=2 \, \Lambda \, \Phi^{1/(D-2)} - \bar{R} \, \Phi^{-1/(D-2)}$, where $\Lambda$ is the cosmological constant and $\bar{R}$ is Ricci scalar of round unit sphere $S^{D-2}$. For the particular case of $D=3$ which will be discussed in  section \ref{sec:reduction-2d}, $X(\Phi)=2\Lambda \Phi$.

\subsection{Solution phase space}\label{sec:2d-SPS}
{{There are three components of the metric and one scalar field. Two diffeomorphisms can be used to fix two of these functions.}}
{Two independent components of equations of motion \eqref{EOM} can then be used to determine two functions among four. One can solve these equations directly, but it would be easier to use diffeomorphisim to construct solution phase space.}  Let us choose one of the coordinates $x$ such that $\Phi =x$. One can show that a generic solution admits a Killing vector field  $k= \epsilon^{\mu \nu} \partial_\mu  \Phi \partial_\nu$, where $\epsilon^{\mu \nu}$ is the Levi-Civita tensor. 
We can choose the other coordinate $y$ to be affine parameter along the curve generated by Killing vector field. In the $x,y$ coordinate system, therefore,  $k= \partial /\partial {y}$. The most general solution in this coordinate system takes the form
\begin{equation}\label{solution-metric}
\d s^2 = 2 \d x \d y + (\mathcal{X}+m) \d y^2
\end{equation}
where $X= \d{} \mathcal{X}/\d{} \Phi$ and  $m$ is an integration constant and we choose ${\cal X}$ such that ${\cal X}(0)=0$.
The Killing vector field $k$ is time like if $\mathcal{X}+m>0$. 
While $m$ is a constant over spacetime we allow it to  vary over the solution phase space (i.e. $m$ has parametric variations, $\delta m \neq 0$). 

The most general solution phase space can be obtained by setting
\begin{equation}\label{SPS}
   x=\Phi(v,r),\qquad y=\Psi(v,r),
\end{equation}
{with non-zero Jacobian, i.e $\partial_{r}\Psi\partial_{v}\Phi-\partial_{r}\Phi\partial_{v}\Psi \neq 0$.} 
Starting from metric \eqref{solution-metric} with \eqref{SPS}, we get 
\begin{equation}
    \begin{split}
         \d s^2 = & \partial_v\Psi \left[ 2 \partial_v \Phi + (\mathcal{X}+m) \partial_v\Psi\right] \d v^2 + \partial_r\Psi \left[ 2 \partial_r \Phi + (\mathcal{X}+m) \partial_r\Psi\right] \d r^2 \\
         &+ 2 \left[ \partial_v\Psi \partial_r\Phi + \partial_r\Psi \partial_v\Phi + (\mathcal{X}+m) \partial_v\Psi \partial_r\Psi \right] \d v \d r .\nonumber
    \end{split}
\end{equation}
In our analysis here, we restrict ourselves to a subset of the above metrics obtained by the gauge fixing, 
\begin{equation}
    g_{rr}=0 \, , \qquad g_{vr}=\eta(v) \, ,
\end{equation}
which may be obtained through diffeos,
\begin{equation}\label{Phi-Omega-2d}
    \Phi (v,r)= \Omega + \eta \, \lambda \, r \, , \qquad  \d \Psi =  \frac{\d v}{ \lambda} \, ,
\end{equation}
where $\Omega \, , \lambda$ are functions of $v$. In this case  the metric can be written as
\begin{equation}\label{2d-solution-GF}
    \d s^2 = -V  \d v^2 + 2\eta \d v \d r \,, \qquad \qquad  V =-\frac{1}{\lambda^2} \left( 2\lambda \partial_v \Phi + \mathcal{X} +m\right)  .
\end{equation}
In our analysis we place the boundary at an arbitrary constant $r$ slice 
and restrict ourselves to a region of spacetime where $V\geq 0$, that is a region where the normal vector to the boundary is spacelike or null, or equivalently the boundary is causal.  That is how we choose the $r$ coordinate. We therefore excise the spacetime at the boundary and analyse $2d$ gravity on a part of spacetime bounded by a causal boundary. 
In this gauge the solution is specified by three functions of $v$, $\eta, \lambda, \Omega$ and a parameter $m$. For later convenience, instead of $\eta$ we use function $\Pi$ defined as 
\begin{equation}
    \Pi = \ln{\left(\frac{\eta \lambda}{\Omega}\right)^2}\, .
\end{equation}

One can recover the null boundary solution space  of \cite{Adami:2020ugu} by the extra requirement that $V$ vanishes at the boundary. Recalling \eqref{2d-solution-GF}, this may be achieved by solving $\lambda$ in terms of the other two functions $\Omega, \eta$ and the parameter $m$.


\subsection{Boundary symmetry generators}\label{sec:2-SymGen}

The vector fields preserving the form of metric \eqref{2d-solution-GF} is
\begin{equation}\label{2d-CBS}
    \xi= T(v)\partial_{v}+\left[Z(v){- \frac{r}{2} W(v)}\right]\partial_{r} \, ,
\end{equation}
{where $T(v)$, $Z(v)$ and $W(v)$ are respectively $v$-supertranslation, $r$-supertranslation and superscaling  in $r$-direction.} The Lie bracket of two vectors of the form \eqref{2d-CBS} is
\begin{equation}\label{2d-algebroid}
[\xi(T_{1},Z_{1},W_{1}),\xi(T_{2},Z_{2},W_{2})]=\xi(T_{12},Z_{12},W_{12})
\end{equation}
where
\begin{subequations}
    \begin{align}
        T_{12}=& T_{1}\partial_{v}T_{2}-(1\leftrightarrow 2)\\
        Z_{12}=& T_{1}\partial_{v}Z_{2}{-\frac{1}{2}}Z_{1}W_{2}-(1\leftrightarrow 2)\\
        W_{12}=& T_{1}\partial_{v}W_{2}-(1\leftrightarrow 2).
    \end{align}
\end{subequations}
As we see $v$-supertranslations form a Witt algebra and $r$-supertranslations and superscalings form a Heisenberg algebra.
The above symmetry generators induce the following transformation laws on fields parameterizing the solution space
\begin{equation}
    \begin{split}
        {\delta_\xi \eta} =&  {\partial_v (T \eta ) -\frac{1}{2}\eta W} \\
        \delta_\xi \Omega = & T\partial_v \Omega+\eta\lambda Z \\
        \delta_\xi \Pi = &- W + T\partial_{v}\Pi - 2 e^{\Pi/2} Z\\
        {\delta_\xi \lambda^{-1}} = &  \partial_{v}(\lambda^{-1}T)\\ 
        \delta_{\xi} m= & 0
    \end{split}
\end{equation}
The last variation ensures that $m$ only captures the parametric variation.
The exact Killing vector in this gauge takes the form
\begin{equation}\label{2d-Killing}
    k=   \lambda \partial_v - \frac{1}{\eta}[\partial_v \Omega + r \partial_v ( \eta \lambda)] \partial_r\, .
\end{equation}
As we see the vector $k$ is not among the symmetry generators \eqref{2d-CBS} in the chosen slicing where $T,W,Z$ are taken to be state independent. In particular, one can show that the adjusted Lie bracket \cite{Barnich:2011mi, Compere:2015knw} of $k$ and the symmetry generators $\xi$ vanishes, $[k, \xi]_{\text{\tiny{adj. bracket}}}=0$, see appendix \ref{sec:change of slicing} for the definition of adjusted bracket. 

Compared to the null boundary case studied in \cite{Adami:2020ugu}, we are allowing for the $r$ supertranslations $Z\partial_r$  that generate fluctuations in the one-form $\d r$ direction. These are transformations which generate fluctuations parametrized by $\lambda$.\footnote{The asymptotic symmetry analysis for $2d$ dilaton gravity has been considered in many previous papers, see in particular \cite{Grumiller:2017qao, Grumiller:2021cwg} and references therein. In these cases, as in our case, they allow for the dilaton $\Phi$ also fluctuate along the boundary. We thank Daniel Grumiller for a comment on this point.}

\subsection{Boundary charges,  $2d$ analysis}\label{sec:2d-bdryCharge}

Having the symmetry generators, in this section we apply the general analysis of appendix \ref{sec:bdryCharge} to the $2d$ gravity case.

\paragraph{Symplectic form.} 
The Lee-Wald symplectic potential for the $2d$ Einstein dilaton gravity is given by
\begin{equation}\label{symp-pot-2dim}
        \Theta^\mu_{\text{\tiny{LW}}}[\delta g,\delta \Phi ;g,\Phi ] = \frac{\sqrt{-g}}{16\pi G} \left[ \Phi \left( \nabla_\nu h^{\mu \nu} - \nabla^\mu h \right) + h \partial^\mu \Phi- h^{\mu \nu} \partial_\nu  \Phi\right] 
\end{equation} 
whose component for the given solution are obtained as
\begin{equation}\label{LW-symplectic-potential}
    16\pi G\Theta^{v}_{\text{\tiny{LW}}}=\lambda \delta\eta \, , \qquad 16\pi G\Theta^{r}_{\text{\tiny{LW}}}=\partial_{v}\left[  \delta \Pi \Omega +\delta (\Phi \,  \sqrt{-g})  \epsilon^{vr}  \right]+\frac{\delta m}{\lambda}
    +\delta \left( \partial_v \Phi  +\frac{\Phi X-2(\mathcal{X}+m)}{\lambda}\right).
\end{equation}
For the JT gravity, $2{\cal X}=X\Phi$ and the second term in the total variation term becomes $\Phi$ independent. We fix the $Y$-freedom upon the physical requirement that the symplectic form (and hence the charge variations computed upon that) are independent of the position of the causal boundary, i.e. they are $r$-independent. This physical requirement is well-justified because the boundary is not a special place in the spacetime\footnote{Noting \eqref{Phi-Omega-2d}, $\Omega=\Phi(r=0)$ and therefore, the choice of $\Omega$ is marking $r=0$.} and that in our analysis we are including the $Z$ transformation which generates displacements in $r$.
The expression \eqref{LW-symplectic-potential} suggests the $Y$-term,
\begin{equation}\label{2d-Y-term}
    Y^{\mu\nu}[\delta g, \delta \Phi ;g , \Phi] = \frac{\delta (\Phi \, \sqrt{-g})}{16\pi G} \, \epsilon^{\mu \nu}
\end{equation}
eliminates the $r$ dependence of the symplectic potential. As discussed in the appendix \ref{sec:bdryCharge}, $W$-freedom (and the boundary Lagrangian) do not contribute to the symplectic form. With this choice for $Y$-term,  the symplectic form on the causal boundary is obtained as
\begin{equation}\label{symplectic-form-m}
        \Omega [\delta \varphi , \delta \varphi ; \varphi] = \frac{1}{16 \pi G} \int \d v \, \left[\partial_v \left( \delta \Pi \wedge \delta \Omega\right)+\delta\lambda^{-1}\wedge\delta m
        \right]
\end{equation}
where $\varphi=\{ \Omega \, , \Pi \, , \lambda\, ; m \}$ is the collection of fields appeared in symplectic form. We stress that \eqref{symplectic-form-m} is $r$-independent and hence we need not specify at which $r$ the integral is computed.

\paragraph{Charge variation.} One can compute the charge variation using \eqref{charge-variation}, to obtain
\begin{equation}\label{Charge-Variation-001'}
           \slashed{\delta} Q_\xi 
           =\frac{1}{16\pi G}\left[W\delta\Omega+2Z\ \delta(\Omega \,e^{\Pi/2})+{T}\left(\frac1{\lambda}\delta m-{\partial_{v}\Pi}\delta\Omega+{\partial_{v}\Omega}  {\delta\Pi} \right)\right].
\end{equation}
In presence of a non-vanishing $m$, we have three surface charges. 

\paragraph{Barnich-Troessaert (BT) charge-flux splitting.} 
Using the BT method, we find 
\begin{subequations}\label{2d-BT-charge-flux'}
    \begin{align}
    16\pi G\, Q^{\text{I}}_{\xi}=& {W} \Omega+2Z\, \Omega \, e^{\Pi/2} + T \left(\frac{m}{\lambda}- \Omega{\partial_{v}\Pi}\right),\\
        16\pi G\, {F}_{\xi}=& T \left[{-m \, \delta (\lambda^{-1})}+\partial_{v}(\Omega\delta\Pi)\right]
    \end{align}
\end{subequations}
with a vanishing central extension term, $ K_{\xi_1,\xi_2}=0$. Let us suppose that the integrable part of charges admit a Laurent expansion in power of $v$ and denote the modes of charges associated with the symmetry generators $\xi(-v^{n{+1}},0,0)$, $\xi(0,v^n,0)$ and $\xi(0,0,v^n)$ respectively by  $\bc{T}_n, \bc{Z}_n, \bc{W}_n$, $n\in\mathbb{Z}$, explicitly,
\begin{equation}
    \bc{W}_n={v^n} \Omega(v)\, ,\quad \bc{Z}_n=2{v^n}\, \Omega (v)\, e^{\Pi(v)/2}\, ,\quad \bc{T}_n ={-{v^{n+1}} \, \left(\frac{m}{\lambda(v)}- \Omega(v){\partial_{v}\Pi(v)}\right)} \, .
\end{equation}
The charge algebra is then,
\begin{subequations}\label{2d-Thermodynamic-slicing-algebra}
    \begin{align}
    \{\bc{T}_n, \bc{T}_m\}
    &= {(n-m)}\bc{T}_{m+n}, \label{2d-TSA-a} \\
    \{\bc{T}_n, \bc{Z}_m\}
    &= {-} m\bc{Z}_{n+m}, \label{2d-TSA-b} \\
    \{\bc{T}_n, \bc{W}_m\}
    &= {-} m\bc{W}_{m+n}, \label{2d-TSA-c} \\
    \{ \bc{W}_m, \bc{Z}_n\}
    &= {\frac{1}{2}}\bc{Z}_{m+n}, \label{2d-TSA-d}\\
    \{\bc{Z}_n, \bc{Z}_m\}
    &=\{\bc{W}_n, \bc{W}_m\}
    =0. \label{2d-TSA-e}
     \end{align}
\end{subequations}
The three integrable part of charges and also the flux  ${F}_{\xi}$ we get from the  BT charge-flux separation equation \eqref{2d-BT-charge-flux'}, namely ${\Omega}, 2\Omega e^{\Pi/2}, {m}/{\lambda}-\Omega{\partial_{v}\Pi}$, are composed of three functions in the solution space, $\Omega,\eta, \lambda$ as well as the parameter $m$. Recalling \eqref{2d-BT-charge-flux'}, one may verify that  commutators in \eqref{2d-TSA-d}, \eqref{2d-TSA-e} are satisfied if  
{$\{\Omega(v), \Pi(v)\}=16\pi G$. }

\paragraph{Noether charge.}
The Lee-Wald Noether potential for the Einstein-Dilaton gravity is given by
\begin{equation}\label{Noether-charge-2dim}
    \text{N}^{\mu\nu}_{\text{\tiny{LW}} \, \xi}=-\frac{\sqrt{-g}}{8\pi G}\left[\Phi\nabla^{[\mu}\xi^{\nu]}+2\xi^{[\mu}\nabla^{\nu]}\Phi\right] .
\end{equation}
To compute the Noether charge, we should add the adopted $Y$-term \eqref{2d-Y-term} to the expression for the Noether charge \eqref{Noether-charge}. Doing so, one may see that the expression of the Noether charge and the integrable part of the charge $Q^{\text{I}}_\xi$ obtained in \eqref{2d-BT-charge-flux'} do not match. However, we note that there is still a $W$-freedom in the Noether charge. One can fix this $W$-freedom upon the requirement that the Noether and BT integrable charge to become identical. One can verify that a $W$-term of the form
\begin{equation}\label{W-term-2dim}
    \mathcal{W}^{\mu}=\sqrt{-g}\left[k^{\mu}+\frac{1}{\lambda}(\Phi X -2 \mathcal{X} - m) n^{\mu}\right]
\end{equation}
where $n_\mu \d x^\mu := - \d v$ is a null co-vector field, does the job and yields $\text{N}_{\xi}= \text{N}^{vr}_\xi = Q^{\text{I}}_\xi$. We note that a very closely related discussion and analysis for fixing the $W$ and $Y$ freedoms has also appeared in \cite{Freidel:2021fxf, Ciambelli:2021vnn, Freidel:2021cjp, Ciambelli:2021nmv, Freidel:2021dxw, Speranza:2022lxr}.

\paragraph{Balance  equation.}
The three integrable part of charges discussed above are functions of $v$, and in particular, there is one of them associated with vector $\partial_v$, $T=1, W=Z=0$, which may be viewed as Hamiltonian (generator of translation along $v$ direction). One may then rewrite \eqref{2d-BT-Bracket-01} for $\xi_2=\partial_v$ and generic $\xi_1=\xi$, as follows,\footnote{The charge variation and  all the associated equations are on-shell. However, since we have explicitly solved the equations of motion functions $\Omega, \Pi$ are independent, so for the balance equation usual and on-shell equality are identical.}
\begin{equation}\label{2d-GCCE}
\frac{\d {} }{\d v} Q^{\text{I}}_{\xi}= \delta_{\partial_{v}}Q^{\text{I}}_{\xi} + \{Q^{\text{I}}_{\partial_v}, Q^{\text{I}}_{\xi}\}\ = -F_{\partial_v}(\delta_\xi g ; g).
\end{equation}
This equation tells us how non-conservation of charges is related to the fluxes. Or alternatively, it specifies how charges balance themselves due to the passage of the flux through the boundary. We note that the flux
$16\pi G\, F_{\partial_v}(\delta_\xi g ; g)= \partial_{v}(\Omega\delta_\xi\Pi)-\delta_\xi(m/\lambda)$, is not a `genuine flux' \cite{Adami:2020ugu, Adami:2021nnf}, as in $2d$ theory we are studying there are no propagating modes with nonvanishing flux through the boundary. To understand this `fake news' an analogy with usual Newtonian mechanics can be helpful. In non-inertial frames there are  Coriolis type forces which are analogous to the fake new here. In our case, as we will see in the next subsection, there are integrable slicing which could be viewed as `inertial frame' in our analogy. 


In a similar way one may choose $\xi_2=\partial_r$ with generic $\xi_1=\xi$. This yields  $\frac{\d {} }{\d r} Q^{\text{I}}_{\xi}=\delta_{\partial_{r}}Q^{\text{I}}_{\xi}+Q^{\text{I}}_{[\partial_{r},\xi]}= 0$. This is of course compatible with the fact that the charges are $r$ independent. 

\subsection{Integrable slicing}\label{sec:2Int-Slice}
As reviewed in the introduction, {the} absence of propagating degrees of freedom in the $2d$ gravity signals existence of slicings in which charges are integrable. To this end, consider the slicing
\begin{equation}
  \hat{W} =W-\partial_{v}{\Pi}\ T +2e^{\Pi/2}\ Z  \, , \qquad \hat{Z}= \Omega e^{\Pi/2}\ Z+ \partial_{v}\Omega T \, ,  \qquad \hat{T}=\frac{T}{\lambda} \, ,
\end{equation}
in which we assume that $\hat{W},\hat{Z},\hat{T}$ are field-independent, while $W, Z, T$ are field-dependent in the hatted-slicing. The surface charges in the hatted slicing become integrable:
\begin{equation}\label{2d-charge-integrable}
          16\pi G\, \slashed{\delta} {{Q}}_\xi= \hat{W} \delta\Omega+\hat{Z} \delta\Pi +\hat{T} \delta m.
\end{equation}
The charge variation in this slicing takes the form,
\begin{equation}\label{2d-integrable-variations}
    \begin{split}
        \delta_\xi \Omega =  \hat{Z},\qquad 
        \delta_\xi \Pi =  -\hat{W},\qquad 
        \delta_\xi \lambda^{-1} =   \partial_{v} \hat{T},\qquad 
        \delta_{\xi} m= 0,
    \end{split}
\end{equation}

We note that the Killing vector $k$ \eqref{2d-Killing} in the hatted slicing takes a very simple form: $\hat{T}=1, \hat{W}=\hat{Z}=0$ {and hence $k$ is among symmetry generators in the hatted slicing}. One can directly check that $m$ is the  charge associated with exact Killing vector field $k$,  $ {{Q}(k)}= M=m / (16\pi G)$. Note also that the hatted slicing is not the only integrable slicing and there are many other such slicings. 

The slicing employed in the charge expression \eqref{2d-charge-integrable} besides integrability of the charge  exhibits another notable feature:  $\lambda$ does not appear in the expression of the charge. One may show that this feature is not limited to the above slicing; there is no integrable slicing in which $\lambda$ appears. Despite this fact, $\lambda$ is not a pure gauge, as it appears in the symplectic form over the solution phase space \eqref{symplectic-form-m} and also in the charge variation \eqref{Charge-Variation-001'}. Nevertheless, recalling \eqref{symplectic-form-m}, we  note that the canonical conjugate to $\lambda^{-1}$ is  $m$ which is a number and not a function of $v$.\footnote{We note that our charge expression is exactly the same as the one appeared in section 6 of \cite{Grumiller:2021cwg}. However, the difference is exactly the fact that in our case the phase space is governed by $3$ functions of $v$ and one parameter $m$, whereas in their case by $2$ functions of $v$ and one parameter.}  To understand this point better, we note that the most general solution space of the $2d$ dilaton gravity has four functions in it\footnote{See \cite{Grumiller:2017qao} for a similar study in AdS$_2$ gravity and its asymptotic symmetries.} and the solution space with three functions and a number we discussed here, is a reduction of that phase space where $m$ is reminiscent of the fourth function of $v$ we have fixed. A similar feature appears once we consider the $2d$ case as a reduction of $3d$ theory over a circle, see next section and in particular subsection \ref{sec:reduction-2d}. As another related comment, one may further reduce the phase space by fixing $\lambda$ to a given constant, e.g. $\lambda_0$, then $\lambda_0^{-1}$ is the canonical conjugate to $m$. In this reduced phase space, only the zero mode of $T$, associated with $\partial_v$ remains as a symmetry generator. 

\paragraph{Symmetry algebra.} The symmetry algebra in the integrable slicing is given by
\begin{equation}\label{2d-algebra}
[\xi(\hat{T}_{1},\hat{Z}_{1},\hat{W}_{1}),\xi(\hat{T}_{2},\hat{Z}_{2},\hat{W}_{2})]_{{\text{\tiny{adj. bracket}}}}=\xi(\hat{T}_{12},\hat{Z}_{12},\hat{W}_{12})
\end{equation}
where
\begin{equation}
    \hat{T}_{12}=0, \hspace{1cm} \hat{W}_{12}=0, \hspace{1 cm} \hat{Z}_{12}=0
\end{equation}
\paragraph{Surface charge algebra.} In the integrable slicing and in the absence of flux,  one may read surface charge algebra
as
\begin{equation}
   \delta_{\xi_{2}}Q_{\xi_{1}} =\left\{ Q_{\xi_{1}} , Q_{\xi_{2}}\right\}=Q_{[\xi_{1},\xi_{2}]}+K_{\xi_1,\xi_2}
\end{equation}
where the central extension term is
\begin{equation}
    K_{\xi_1,\xi_2}=\frac{1}{16\pi G}\left(\hat{W}_{1}\hat{Z}_{2}-\hat{Z}_{1}\hat{W}_{2}\right).
\end{equation}
The above charge algebra implies
\begin{equation}\label{2d-Int-Charge-algebra}
      \{ \Omega (v), \Pi (v)\}= 16\pi G.
\end{equation}
The mass parameter $m$  commutes with all charges, as $[k, \xi]_{\text{\tiny{adj. bracket}}}=0$. This is in accord with  general arguments in \cite{Hajian:2015xlp} that all exact symmetry (Killing) charges commute with charges associated with non-trivial diffeomorphisms.

\section{Causal Boundary Symmetries, $3d$ Case}\label{sec:3''}

In this section we study the causal boundary charges for the $3d$ Einstein-$\Lambda$ theory, described by the action, 
\begin{equation}\label{3D-action}
    S= \frac{1}{16\pi G} \int \d{}^3 x \sqrt{-g}  \left( R - 2\Lambda \right)
\end{equation}
where ${R}$ is Ricci scalar and $\Lambda$ is the cosmological constant. In our analysis we do not fix the sign of $\Lambda$, $\Lambda  \lesseqgtr 0$. The field equations for this action are 
\begin{equation}\label{EOM-3d}
R_{\mu\nu}= -2\Lambda g_{\mu\nu}.
\end{equation}

\subsection{Causal surface solution phase space}\label{sec:review}

We adopt Gaussian null-type coordinate system in which $v,r,\phi$ are respectively advanced time, radial and angular coordinates. The three dimensional line-element in Gaussian null-type coordinate system is \cite{Adami:2020amw,Adami:2020ugu,Adami:2021nnf}
\begin{equation}\label{G-N-T-metric''}
    \d s^2=  -V \d v^2 + 2 \eta \d v \d r + {\cal R}^2 \left( \d \phi + U \d v \right)^2\, .
\end{equation}
where $V,\mathcal{R}, U$ are generic functions on spacetime and $\eta$ depends only on $v$ and $\phi$. We also assume $\phi \sim \phi+2\pi$ and that all metric components are periodic functions of  $\phi$. While {$v$} generically ranges in $\mathbb{R}$, we restrict the $r$ coordinate to be larger {than} an arbitrary value $r$ at which we place our boundary, see Fig. \ref{fig:causal-boundary}. We assume that in the half of the spacetime we consider $V\geq 0$, 
so that $\partial_v$ is a causal vector and $\partial_r$ is a null vector, and without loss of generality we take $\Omega, \eta,\lambda>0$. 

{The gauge  we adopt here to describe solution space is different from the ones that has been widely used in the literature for the asymptotic symmetry analysis. Bondi gauge introduced in seminal works \cite{Bondi:1962,Sachs:1962wk,Sachs:1962zza} leads to the Bondi-van der Burg-Metzner-Sachs (BMS) algebra as the asymptotic symmetries of asymptotically flat spacetimes. The BMS algebra is enlarged in \cite{Barnich:2010eb,Barnich:2011mi} to include superrotations. Other extensions of BMS or Virasoro algebras can be found in \cite{Campiglia:2020qvc,Compere:2020lrt, Ruzziconi:2020wrb, Geiller:2021vpg, Alessio:2020ioh}.}

Equations of motion \eqref{EOM-3d} specify the $r$ dependence of fields as \cite{Adami:2020ugu}
\begin{subequations}\label{metric-components''}
    \begin{align}
    {\cal R}=&\Omega + \lambda \, \eta \, r  \label{calR-3d}\\
    U=&{\cal U} +  \frac{1}{\lambda \, {\cal R}}\, \frac{\partial_\phi \eta}{ \eta }  + \frac{\Upsilon}{2 \lambda {\cal R}^2} \\
         V= &\frac{1}{\lambda^2}\left( - \Lambda \mathcal{R}^2 -\mathcal{M}  + \frac{\Upsilon^2}{4  {\cal R}^2} - \frac{2\mathcal{R}}{\eta }   \mathcal{D}_v ( \eta \lambda ) +\frac{\Upsilon}{\mathcal{R}}\, \frac{\partial_\phi \eta}{ \eta } \right)
    \end{align}
\end{subequations}
where $\Omega, \lambda, \eta, \Upsilon, \mathcal{U}, \mathcal{M}$ are  functions of $v, \phi$. Moreover, \eqref{G-N-T-metric''} with \eqref{metric-components''} solve Einstein field equations if,
\begin{subequations}\label{M-Upsilon-EoM''}
\begin{align}
    &\mathcal{E}_{\hat{\mathcal{M}}}:=\mathcal{D}_{v}\hat{\mathcal{M}}+\Lambda\lambda\partial_{\phi}\left(\frac{\hat{\Upsilon}}{\lambda^2}\right)+2\partial_{\phi}^3\mathcal{U}=0 \label{EOM-M}\\
    &\mathcal{E}_{\hat{\Upsilon}}:=\mathcal{D}_{v}\hat{\Upsilon}-\lambda\partial_{\phi}\left(\frac{\hat{\mathcal{M}}}{\lambda^2}\right)+2\partial_{\phi}^3(\lambda^{-1})=0 \label{EOM-Upsilon}
\end{align}
\end{subequations}
where
\begin{equation}\label{hatM-hatUsp''}
    \begin{split}
       &  \hat{\Upsilon}={\Upsilon+\Omega\partial_{\phi}\Pi}\, , \qquad {\Pi :=\ln \left( \frac{\eta \lambda}{\Omega} \right)^2}\, ,\\
       & {\hat{\mathcal{M}}= \mathcal{M} + {\lambda\Omega\mathcal{D}_{v}\Pi}+ \left(\frac{\partial_{\phi}\eta}{\eta}\right)^2+3\left(\frac{\partial_{\phi}\lambda}{\lambda}\right)^2-2\frac{\partial_{\phi}^2\lambda}{\lambda}}
    \end{split}
\end{equation}
and the differential operators $\mathcal{D}_v$ and $\mathcal{L}_\mathcal{U}$ which act on a codimension one function $O_w (v,\phi)$ of weight $w$ is defined through
\begin{subequations}
    \begin{align}
    \mathcal{D}_v O_w= &\partial_v O_w - \mathcal{L}_\mathcal{U} O_w \, ,\\
    \mathcal{L}_{\mathcal{U}} O_w= & \mathcal{U} \partial_\phi O_w + w O_w \partial_\phi \mathcal{U} \, , 
    \end{align}
\end{subequations}
where $\mathcal{U}$ is a function of weight $-1$. 
Weights of  different functions can be found in Table \ref{Table-1}.  
\begin{table}[h]
\centering
\begin{tabular}{ |l|l| }
  \hline
  $w= -1$  & $\mathcal{U}$  ,  $Y$\\
  $w= 0$  & $\eta$  , $T$ , $W$ , $Z$ , $\Pi$ , $\partial_{v}$\\
  $w= 1$  & $\Omega$  , $\lambda$ , $\partial_{\phi}$ \\
  $w= 2$  & $\hat{\mathcal{M}}$ , $\hat{\Upsilon}$ \\
  \hline
\end{tabular}
\caption{Weight $w$ for various quantities defined and used in this section.}\label{Table-1}
\label{table:weight}
\end{table}
We remark that with the above conventions, ${\cal R}= {\Omega(1+e^{\Pi/2} r)}$ and  the third order $\phi$ derivative terms in \eqref{M-Upsilon-EoM''}  come from  substitution of other field equations to simplify these equations. We also note that if $f:=\int_\phi \lambda$ then  the last two terms are $-2 S(f, \phi)$, where $S$ is the Schwarzian derivative.
One can treat these equations as equations for $\mathcal{U}$ and $\lambda$ which only involve $\phi$ (and not $v$) derivatives of these functions. Finally,  up to some functions of only $v$,   the solution space is spanned by four codimension one functions $\mathcal{M}, \Upsilon, \Omega, \Pi$.

\subsection{Causal boundary symmetry} 

Let the boundary  be an arbitrary constant $r$ surface, $\cc_r$. $\cc_r$ is a causal (timelike or null)  surface and we are interested in formulating physics in  one side of the boundary which contains $r\to \infty$, see Fig. \ref{fig:causal-boundary}. The vector field
\begin{equation}\label{null-bondary-sym-gen''}
    \xi=T\partial_{v}+\left[Z  - \frac{r }{2}\, W -\frac{\Upsilon}{2\eta\lambda^2\mathcal{R}} \, \partial_{\phi}T - \frac{1}{\eta^2\lambda }\partial_{\phi}\left(\frac{\eta\partial_{\phi}T}{\lambda}\right)\right]\partial_{r}+\left(Y+\frac{\partial_{\phi}T}{\lambda\mathcal{R}}\right)\partial_{\phi}
\end{equation}
preserves the form of metric \eqref{G-N-T-metric''} and hence moves us in the solution space constructed above. {This vector field is parametrized by supertranslation in $v$-direction $T(v,\phi)$, supertranslation in $r$-direction $Z(v,\phi)$, superscaling $W(v,\phi)$, and superrotation $Y(v,\phi)$.} Under the action of $\xi$, functions in the metric have the following variations 
\begin{subequations}\label{delta-fields}
    \begin{align}
       \delta_\xi \eta =&  \mathcal{D}_v (T \eta )  + \hat{Y} \partial_\phi \eta -\frac{1}{2}\eta W \\
    \delta_{\xi}\lambda =& T\mathcal{D}_{v}\lambda-\lambda\mathcal{D}_{v}T +\partial_{\phi}(\lambda \hat{Y})\\
   \delta_{\xi}\mathcal{U} =& \mathcal{D}_{v}\hat{Y} +\frac{\Lambda\partial_{\phi}T}{\lambda^2}\\
      \delta_{\xi}\Omega =& T \, \mathcal{D}_v \Omega+\partial_{\phi}(\Omega \hat{Y})+\eta\lambda Z \\
      \delta_{\xi}\hat{\Upsilon} \approx & \hat{T}\partial_{\phi}\hat{\mathcal{M}}+2\hat{\mathcal{M}}\partial_{\phi}\hat{T}+\hat{Y}\partial_{\phi}\hat{\Upsilon}+2\hat{\Upsilon}\partial_{\phi}\hat{Y}-2\partial_{\phi}^{3}\hat{T}\\
   \delta_{\xi}\hat{\mathcal{M}} \approx & \hat{Y}\partial_{\phi}\hat{\mathcal{M}}+2\hat{\mathcal{M}}\partial_{\phi}\hat{Y}-\Lambda(\hat{T}\partial_{\phi}\hat{\Upsilon}+2\hat{\Upsilon}\partial_{\phi}\hat{T})-2\partial_{\phi}^{3}\hat{Y} \\
   \delta_{\xi}\Pi=& - W +  T \mathcal{D}_v \Pi {-2}e^{\Pi/2} Z +\hat{Y}\partial_{\phi}\Pi 
   \end{align}
\end{subequations} 
where $\hat{Y}= Y+ \mathcal{U} T$ and $\hat{T}=T/\lambda$ and $\approx$ indicates on-shell equality in which equations of motion \eqref{M-Upsilon-EoM''} are  used.

\paragraph{Causal Boundary Symmetry Algebra.} 
Since the causal boundary symmetry generators \eqref{null-bondary-sym-gen''} depend on the field in solution space to compute their solution phase space Lie bracket we need to adjust for the field variations \cite{Barnich:2011mi, Compere:2015knw} {(see appendix \ref{sec:change of slicing} for more details)}. Using the adjusted Lie bracket we have
\begin{equation}\label{3d-NBS-KV-algebra''}
    [\xi(  T_1, Z_1, W_1, Y_1), \xi( T_2, Z_2,  W_2, Y_2)]_{_{\text{adj. bracket}}}=\xi(  T_{12}, Z_{12}, W_{12}, Y_{12})
\end{equation}
where 
\begin{subequations}\label{W12-T12-Z12-Y12''}
\begin{align}
    &T_{12}=(T_{1}\partial_{v}+Y_{1}\partial_{\phi})T_{2}-(1\leftrightarrow 2)\\
    &Z_{12}=(T_{1}\partial_{v}+Y_{1}\partial_{\phi})Z_{2}+\frac{1}{2}W_{1}Z_{2}-(1\leftrightarrow 2)\, \\
    &W_{12}=(T_{1}\partial_{v}+Y_{1}\partial_{\phi})W_{2}-(1\leftrightarrow 2)\\
    &Y_{12}=(T_{1}\partial_{v}+Y_{1}\partial_{\phi})Y_{2}-(1\leftrightarrow 2)
\end{align}
\end{subequations}
One may wonder how we have fixed the quite non-trivial field dependence in the vector fields \eqref{null-bondary-sym-gen''}. In fact, there are (infinitely) many other choices for symmetry generators and their field dependence which rotate us within the solution space. The specific form \eqref{null-bondary-sym-gen''} has the feature that it leads to an algebra; there is no field dependence in \eqref{W12-T12-Z12-Y12''}. Other choices typically lead to  algebroids, where the expression in $T_{12}, Z_{12}, W_{12}, Y_{12}$ are field dependent. One should also note that the specific field dependence in \eqref{null-bondary-sym-gen''}, while special, is not the only one which yields the algebra (and not algebriod) structure.  

\subsection{Surface charge analysis} \label{sec:3.3}
Consider the the Einstein-Hilbert Lagrangian $L_{0} =L_{\text{\tiny EH}}[g]$. The Lee-Wald symplectic potential \cite{Lee:1990nz} can be read as
\begin{equation}\label{Theta-Y''}
    \Theta^{\mu}_{_{\text{\tiny{LW}}}} [g; \delta g]=\frac{\sqrt{-g}}{8 \pi G} \nabla^{[\alpha} \left( g^{\mu ] \beta} \delta g_{\alpha \beta} \right)\,,
\end{equation}
For the Einstein gravity theory
\begin{equation}
    {\mathcal{Q}_{_{\text{\tiny{LW}}} \,\xi}^{\mu \nu}} =\frac{\sqrt{-g}}{8 \pi G}\, \Big( h^{\lambda [ \mu} \nabla _{\lambda} \xi^{\nu]} - \xi^{\lambda} \nabla^{[\mu} h^{\nu]}_{\lambda} - \frac{1}{2} h \nabla ^{[\mu} \xi^{\nu]} + \xi^{[\mu} \nabla _{\lambda} h^{\nu] \lambda} - \xi^{[\mu} \nabla^{\nu]}h \Big)
\end{equation}
is the ordinary Lee-Wald charge variation {density}. We are interested in surface charges computed on a codimension 2 transverse surface $\Sigma$ with binormal $\epsilon^{\mu \nu}$. The surface element hence is $\d S_{\mu\nu} = \epsilon_{\mu\nu}  \mathcal{R} \d{} \phi $. The surface charge variation on transverse surface can be defined as $\slashed{\delta} Q_\xi := \int_{\Sigma} Q^{\mu\nu}_\xi \d S_{\mu\nu}$ where $Q^{\mu\nu}_\xi= \mathcal{Q}_\xi^{\mu \nu} / \sqrt{-g} $ is a skew-symmetric tensor. 

In our case $\Sigma$ may be taken a constant $v$ slice on the boundary ${\cal C}_r$. Therefore, in our case $\Sigma$ is a constant $v,r$ surface $\cc_{r,v}$. The Lee-Wald surface charge variation reads
\begin{equation}\label{surface-charge-LW''}
       \hspace{-0.3 cm}     \slashed{\delta} Q_{\text{\tiny{LW}}}(\xi) = {\frac{1}{16\pi G}}\oint_{\cc_{r,v}} \d \phi \left(
            -\delta_{\xi}\Pi \delta\Omega +\delta_{\xi}\Omega  \delta \Pi + (Y+\mathcal{U}T)  \delta\hat{\Upsilon}  + \frac{T}{\lambda}  \delta\hat{\mathcal{M}} +\delta_{\xi}n^r\delta\sqrt{-g}-\delta n^r\delta_{\xi}\sqrt{-g}\right).
\end{equation}
The last two terms in the Lee-Wald surface charge variation \eqref{surface-charge-LW''} depend on the radial coordinate and they diverge at infinity.  As in the $2d$ case,  we fix the $Y$-freedom such that the symplectic form and surface charges do not depend on the arbitrary $r$ at which the causal boundary resides.  This can be achieved  by adding the covariant $Y$-term,  
\begin{equation}\label{Y-term-3dim}
    Y^{\mu\nu}[ g; \delta g]= \frac{ \delta \sqrt{-g}}{8 \pi G} \, \epsilon^{\mu \nu} \, .
\end{equation}
Upon addition of this $Y$-term,  surface charge variation in the slicing defined by $\delta T= \delta Y= \delta W =\delta Z=0$ becomes $r$-independent and takes the form
\begin{equation}\label{surface-charge-Thermo''}
    \hspace{-0.25 cm}   {\slashed{\delta} Q_\xi = {\frac{1}{16\pi G}}\oint_{\cc_{r,v}} \d \phi \left[ 
            W \delta\Omega + 2Z \delta(\Omega \, e^{\Pi/2}) +Y\delta \Upsilon+ T \left( -\mathcal{D}_{v}\Pi \, \delta\Omega+{\mathcal{U}} \delta\Upsilon+\mathcal{D}_{v}\Omega\ \delta\Pi+\lambda^{-1}\delta{\hat{\mathcal{M}}}\right)
            \right].}
\end{equation}

\paragraph{BT integrable-flux splitting.} The charge variation \eqref{surface-charge-Thermo''} in the adopted slicing is clearly non-integrable. 
One may  use the adjusted bracket method of Barnich and Troessaert \cite{Barnich:2011mi} to separate integrable $Q^{\text{I}}_\xi$ and flux $ F_{\xi}(\delta g; g)$ parts, see appendix \ref{sec:bdryCharge}. 
Applying the adjusted bracket to the charge variation \eqref{surface-charge-Thermo''} yields,
\begin{subequations}\label{3d-BT}
    \begin{align}
      &  Q^{\text{I}}_{\xi}=  \frac{1}{16\pi G}\int \d \phi \left[  W  \, \Omega +2 Z \, \Omega\, e^{\Pi/2} + Y \, \Upsilon +  T \bigl(\-\mathcal{D}_{v}\Pi \Omega + {\mathcal{U}}\, \Upsilon +\lambda^{-1}\ \hat{{\mathcal{M}}}\bigr)    \right] \, , \label{3d-BT-charge-thermod} \\
      &  F_{\xi}(\delta g; g)=\frac{1}{16\pi G}\int \d \phi \, T \left[  -\hat{\mathcal{M}} \delta (\lambda^{-1} )-   {\Upsilon}\,\delta {\mathcal{U}} {+\Omega\delta\mathcal{D}_{v}\Pi} +\delta\Pi \mathcal{D}_{v}\Omega\right] \, . \label{3d-BT-flux-thermod}
    \end{align}
\end{subequations}
The central term reads as
\begin{equation}\label{3d-BT-central-charge-thermod}
        K_{\xi_1,\xi_2}= \frac{1}{8\pi G}\int \d \phi  \, \frac{1}{\lambda}\left[{T}_{2}\partial_{\phi}^3 ({Y}_{1}+\mathcal{U}T_1)-{T}_{1}\partial_{\phi}^3 ({Y}_{2}+\mathcal{U} T_2)\right].
\end{equation}
While with the field dependence of \eqref{null-bondary-sym-gen''} has been chosen such that the adjusted bracket of symmetry generators \eqref{3d-NBS-KV-algebra''} are field independent, the ``central term'' $ K_{\xi_1,\xi_2}$ \eqref{3d-BT-central-charge-thermod} is field-dependent, as it depends on $\lambda$ and $\mathcal{U}$. Therefore, the BT bracket of charges form an algebroid, rather than an algebra. Despite of this fact, as we will see, one may use this slicing for the balance equation. 

The above charge-flux separation works for generic values of $\Lambda$. For a vanishing cosmological constant $\Lambda=0$, however, one can use the $\mathcal{A}$-freedom, \emph{cf}. Appendix \ref{sec:bdryCharge}, to remove the field dependence of the central term. In particular, for 
\begin{subequations}
    \begin{align}
      &  Q^{\text{I}}_{\xi}=  \frac{1}{16\pi G}\int \d \phi \left[  W  \, \Omega +2 Z \, \Omega\, e^{\Pi/2} + Y \, \Upsilon +  T \bigl(\-\mathcal{D}_{v}\Pi \Omega + {\mathcal{U}}\, \Upsilon\bigr)    \right] \, , \label{3d-BT-charge-thermod} \\
      &  F_{\xi}(\delta g; g)=\frac{1}{16\pi G}\int \d \phi \, T \left[  \lambda^{-1}  \delta\hat{\mathcal{M}}-   {\Upsilon}\,\delta {\mathcal{U}} {+\Omega\delta\mathcal{D}_{v}\Pi} +\delta\Pi \mathcal{D}_{v}\Omega\right] \, . \label{3d-BT-flux-thermod}
    \end{align}
\end{subequations}
we get $K_{\xi_1,\xi_2}=0$ for $\Lambda=0$. Let us assume that the integrable part of charges admit a Laurent expansion in power of $v$ and perform a Fourier transformation in $\phi$ and denote the modes of charges associated with the symmetry generators $\xi(-v^{n{+1}}e^{im\phi},0,0,0)$, $\xi(0,v^n e^{im\phi},0,0)$, $\xi(0,0,v^n e^{im\phi},0)$ and $\xi(0,0,0,i v^n e^{im\phi})$ respectively by  $\bc{T}_{n,m}, \bc{Z}_{n,m}, \bc{W}_{n,m}, \bc{Y}_{n,m}$, $n,m\in\mathbb{Z}$, explicitly,
\begin{equation}
    \begin{split}
        &\bc{W}_{n,m}={\frac{1}{2\pi}\int_{0}^{2\pi} \d \phi \,} {v^n e^{im\phi}} \Omega(v,\phi),\hspace{1 cm} \bc{Z}_{n,m}={\frac{1}{2\pi}\int_{0}^{2\pi} \d \phi \,} {v^n e^{im\phi}}\, \Omega (v,\phi)\, e^{\Pi(v,\phi)/2},\\
        &\bc{Y}_{n,m}={\frac{1}{2\pi}\int_{0}^{2\pi} \d \phi \,} i{v^n e^{im\phi}}\Upsilon(v,\phi) ,\quad \bc{T}_{n,m} =-{\frac{1}{2\pi}\int_{0}^{2\pi} \d \phi \,} {v^{n+1} e^{im\phi}} \,\bigl(\-\mathcal{D}_{v}\Pi(v,\phi) \Omega(v,\phi) + \mathcal{U}(v,\phi)\, \Upsilon(v,\phi)\bigr) \, .
    \end{split}
\end{equation}
The charge algebra yields,
\begin{subequations}\label{3d-Thermodynamic-slicing-algebra}
    \begin{align}
    \{\bc{T}_{n,m}, \bc{T}_{k,l}\}_{_{\text{MB}}} 
    &= (n-k)\bc{T}_{n+k,m+l} \, , \hspace{1 cm} \{\bc{Y}_{n,m}\, , \bc{Y}_{k,l}\}_{_{\text{MB}}}=(m-l)\bc{Y}_{n+k,m+l}\, , \\
    \{\bc{T}_{n,m}, \bc{Z}_{k,l}\}_{_{\text{MB}}}
    &= -k \bc{Z}_{n+k,m+l} \, ,  \hspace{1.7 cm} \{\bc{Y}_{n,m}\, , \bc{Z}_{k,l}\}_{_{\text{MB}}}= -l \bc{Z}_{n+k,m+l} \,  ,\\
    \{\bc{T}_{n,m}, \bc{W}_{k,l}\}_{_{\text{MB}}} 
    &= -k \bc{W}_{n+k,m+l}\, ,  \hspace{1.5 cm} \{\bc{Y}_{n,m} \, , \bc{W}_{k,l}\}_{_{\text{MB}}}= -l \bc{W}_{n+k,m+l} \, ,\\
    \{ \bc{W}_{n,m}, \bc{Z}_{k,l}\}_{_{\text{MB}}} 
    &=\frac{1}{2} \bc{Z}_{n+k,m+l}\, ,  \hspace{2 cm} \{ \bc{Y}_{n,m}\, , \bc{T}_{k,l}\}_{_{\text{MB}}}= l \bc{T}_{n+k,m+l}+n \bc{Y}_{n+k,m+l} \, .
     \end{align}
\end{subequations}
The $\bc{T}_{n,m}, \bc{Y}_{n,m}$ part of the algebra is  diffeomorphisms on the $2d$ cylinder spanned by $v,\phi$, ${\cal A}_{C_2}$ and $\bc{Z}_{k,l}, \bc{W}_{k,l}$ are in vector representation of the ${\cal A}_{C_2}$ algebra. 
This result  coincides with the fact that in the flat limit of two Virasoro algebras at the Brown-Henneaux central charge, we lose one of central charges \cite{Krishnan:2013wta}.
\paragraph{Noether charge.} The Lee-Wald contribution to the  Noether charge, which is nothing but the Komar charge density, is given as
\begin{equation}
    \text{N}_{\text{\tiny{LW}}\,\xi}^{\mu \nu}= -\frac{\sqrt{-g}}{8\pi G}\, \nabla^{[\mu}\xi^{\nu]}\,.
\end{equation}
As discussed, the Noether charge receives contributions from both $W$ and $Y$ terms. By fixing the $Y$-freedom as in  \eqref{Y-term-3dim}, and  $W$-freedom as
\begin{equation}\label{W-term-3dim}
    16\pi G\, \mathcal{W}^{v}={3\partial_r\mathcal{R}} \, , \qquad 16\pi G \, \mathcal{W}^{r}=-\frac{\hat{\mathcal{M}}}{\lambda}-\frac{2\Lambda}{\lambda}\mathcal{R}^{2}-3\mathcal{D}_{v}\mathcal{R}{+}3\partial_{\phi}\left(\mathcal{R} U -\mathcal{R}\mathcal{U}\right),\qquad  \mathcal{W}^{\phi}=0 \, ,
\end{equation}
the Noether charge becomes identical to the integrable part of charge \eqref{3d-BT-charge-thermod}.

\paragraph{Balance equation.} The BT charge-flux splitting \eqref{3d-BT} written for $\xi_1=\partial_v$ and generic $\xi_2$ yields the balance equation:
\begin{equation}\label{GCCE-v-3d-1}
    \frac{\d {}}{\d v} Q^{\text{I}}_{\xi} = \delta_{\partial_{v}}Q^{\text{I}}_{\xi}+Q^{\text{I}}_{[\partial_{v},\xi]} \approx -F_{\partial_{v}}(\delta_{\xi}g; g)+K_{\xi,\partial_{v}}.
\end{equation}
Let us denote the charges as
$$
Q^{\text{I}}_{_{\xi}} := 
\int \d{} \phi \ \mathcal{Q}^{\text{I}}_{\xi}.
$$
Considering time derivative of surface charges associated with $\xi=\xi(T,0,0,0)$ and $\xi=\xi(0,Y,0,0)$ we find
\begin{subequations}\label{GCCE-T-Y}
    \begin{align}
        \frac{\d {}}{\d v}Q^{\text{I}}_{_{T}}= &{\int \d{} \phi \,\mathcal{D}_{v} \mathcal{Q}^{\text{I}}_{{_T}} - \frac{1}{16\pi G}\int \d{} \phi \,T\big(\mathcal{U}\mathcal{E}_{\hat{\Upsilon}}+\lambda^{-1}\mathcal{E}_{\hat{\mathcal{M}}}\big) }\\
        \frac{\d {}}{\d v}Q^{\text{I}}_{_{Y}}= &{\int \d \phi \, \mathcal{D}_{v}\mathcal{Q}^{\text{I}}_{_{Y}}-\frac{1}{16\pi G}\int \d{} \phi\, Y\mathcal{E}_{\hat{\Upsilon}}}.
    \end{align}
\end{subequations}
and for  $\xi=\xi(0,0,Z,0)$ and $\xi=\xi(0,0,0,W)$ the balance equation yields two identities
\begin{equation}\label{GCCE-Z-W}
        \frac{\d {}}{\d v}Q^{\text{I}}_{_{Z}}=\int \d \phi\,\ \mathcal{D}_{v}\mathcal{Q}^{\text{I}}_{_{Z}},\qquad 
        \frac{\d {}}{\d v}Q^{\text{I}}_{_{W}}= \int \d \phi\,\ \mathcal{D}_{v} \mathcal{Q}^{\text{I}}_{_{W}},
\end{equation}
The above, especially \eqref{GCCE-T-Y}, makes it clear that the balance equation is a manifestation of  the equations of motion projected and computed at the boundary, $\mathcal{E}_{\hat{\Upsilon}}=0$ $\mathcal{E}_{\hat{\mathcal{M}}}=0$.

Thanks to the $Y$-freedom, we also have a simple balance equation in the radial direction
\begin{equation}
    \frac{\d {}}{\d r} Q^{\text{I}}_{\xi}=\delta_{\partial_{r}}Q^{\text{I}}_{\xi}+Q^{\text{I}}_{[\partial_{r},\xi]}=0.
\end{equation}
The above is a consequence of $r$-independence of surface charges, as we had in the $2d$ analysis.
\subsection{Surface charges in integrable Heisenberg slicing}\label{sec:3d-int-slicing} 

There is no bulk propagating degree of freedom in $3d$ gravity and hence there should exist slicings in which the surface charge variation is integrable. To see this, consider the change of slicing,
\begin{equation}\label{hat-slicing''}
    \hat{Z} =\delta_{\xi}\Omega \, , \qquad \hat{W}=-\delta_{\xi}\Pi \, , \qquad \hat{Y}= Y+\mathcal{U} T \, , \qquad \hat{T}= \frac{T}{\lambda} \, ,
\end{equation}
in which the surface charge variation takes the form
\begin{equation}\label{surface-charge-Y-infty''}
            {\delta} Q_\xi :=\frac{1}{16\pi G} \oint_{\cc_{r,v}} \d \phi \left(
            \hat{W}\delta\Omega +\hat{Z} \delta \Pi+\hat{Y} \delta\hat{\Upsilon}  + \hat{T}\delta\hat{\mathcal{M}} \right).
\end{equation}
The above is manifestly integrable if we take the new symmetry generators \eqref{hat-slicing''} to be field independent, i.e. 
if we assume $\delta \hat Z=\delta \hat Y=\delta \hat W=\delta \hat T=0$. 

\paragraph{Causal Boundary Symmetry Algebra.} 
 Using the adjusted Lie bracket we have
\begin{equation}\label{3d-NBS-KV-algebra}
    [\xi(  \hat{T}_1, \hat{Z}_1, \hat{W}_1, \hat{Y}_1), \xi( \hat{T}_2, \hat{Z}_2,  \hat{W}_2, \hat{Y}_2)]_{_{\text{adj. bracket}}}=\xi(  \hat{T}_{12}, \hat{Z}_{12}, \hat{W}_{12}, \hat{Y}_{12})
\end{equation}
where 
\begin{subequations}\label{W12-T12-Z12-Y12}
\begin{align}
    &\hat{T}_{12}=\hat{T}_{1}\partial_{\phi}\hat{Y}_{2}-\hat{Y}_{2}\partial_{\phi}\hat{T}_{1}-(1\leftrightarrow 2)\\
    &\hat{Y}_{12}=\hat{Y}_{1}\partial_{\phi}\hat{Y}_{2}-\Lambda\hat{T}_{1}\partial_{\phi}\hat{T}_{2}-(1\leftrightarrow 2)\\
    &\hat{Z}_{12}=0\, \\
    &\hat{W}_{12}=0\,
\end{align}
\end{subequations}
\paragraph{Surface Charge Algebra.} 
The charge algebra in the Heisenberg slicing is
\begin{equation}\label{3d-BT-Bracket-02''}
    \left\{Q^{\text{I}}_{\xi_{1}},Q^{\text{I}}_{\xi_{2}}\right\}_{{{\text{\tiny{MB}}}}} =\, Q^{\text{I}}_{[\xi_{1},\xi_{2}]_{{\text{adj. bracket}}}}+K_{\xi_1,\xi_2}
\end{equation}
where
\begin{equation}
    K_{\xi_1,\xi_2}=\frac{1}{8\pi G} \oint_{\cc_{r,v}} \d \phi \, (\hat{T}_{2}\partial_{\phi}^3 \hat{Y}_{1}-\hat{T}_{1}\partial_{\phi}^3 \hat{Y}_{2})+\frac{1}{16\pi G} \oint_{\cc_{r,v}} \d \phi \, (\hat{Z}_{2}\hat{W}_{1}-\hat{Z}_{1}\hat{W}_{2})
\end{equation}
The explicit form of the algebra is given by
\begin{subequations}\label{Heisenberg-direct-sum-algebra''}
    \begin{align}
        &\{\Omega(v,\phi),\Pi(v,\phi')\}=16\pi G\ \delta\left(\phi-\phi'\right)\\
        &\{\hat{\Upsilon}(v,\phi),\hat{\Upsilon}(v,\phi')\}=16\pi G\left(\hat{\Upsilon}(v,\phi')\partial_{\phi}-\hat{\Upsilon}(v,\phi)\partial_{\phi'}\right)\delta\left(\phi-\phi'\right)\,\\
        &\{\hat{\mathcal{M}}(v,\phi),\hat{\mathcal{M}}(v,\phi')\}=-16\pi G\Lambda\left(\hat{\Upsilon}(v,\phi')\partial_{\phi}-\hat{\Upsilon}(v,\phi)\partial_{\phi'}\right)\delta\left(\phi-\phi'\right)\,\,\\
        &\{\hat{\Upsilon}(v,\phi),\hat{\mathcal{M}}(v,\phi')\}=16\pi G\left(\hat{\mathcal{M}}(v,\phi')\partial_{\phi}-\hat{\mathcal{M}}(v,\phi)\partial_{\phi'}-2\partial_{\phi}^3\right)\delta\left(\phi-\phi'\right)\,
    \end{align}
\end{subequations}
Brackets not displayed vanish.  {For flat case, $\Lambda =0$, the algebra \eqref{Heisenberg-direct-sum-algebra''} is direct sum of the Heisenberg and the $\mathfrak{bms}_3$ algebra. The main deference between the $\mathfrak{bms}_3$ subalgebra and the $\mathfrak{bms}_3$ algebra obtained as the symmetry structure of asymptotically flat spacetimes in $3d$ \cite{Barnich:2006av,Barnich:2012aw} is that the charges  here have explicit  $v$ dependence while the structure constants are still $v$ independent. Put differently, for any constant $v$ slice we find the same algebra Heisenberg $\oplus\ \mathfrak{bms}_3$ algebra.} For the $\Lambda<0$, for any constant $v$ slice, we get a direct sum of Heisenberg and two copies of Virasoro algebras at Brown-Henneaux central charge (see below). This matches the result found in \cite{Geiller:2021vpg} at asymptotic infinity for AdS$_3$ case. For the $\Lambda>0$, at any constant $v$ slice, we get a direct sum of Heisenberg and the algebra obtained in \cite{Compere:2014cna}.
\paragraph{Pre-symplectic form.} 
The $r$--component of symplectic potential \eqref{Theta-Y''} for our solution space \eqref{G-N-T-metric''} is 
\begin{equation}
\begin{split}
    16\pi G\Theta^r[\delta g, g]=&\lambda^{-1}\delta\hat{\mathcal{M}}-\hat{\Upsilon}\, \delta\mathcal{U}+\partial_{v}(\Omega\delta\Pi) + \delta\left[\frac{2\Lambda}{\lambda}\mathcal{R}^2+3\partial_{v}\mathcal{R}\right]\\
    &+\partial_{\phi}\Biggl\{\frac{1}{\lambda}\partial_{\phi}\left(\frac{2\delta\lambda}{\lambda}\right)-\mathcal{R}\left[3\delta\mathcal{U}+\mathcal{U}\frac{\delta(\eta\lambda^2)}{\eta\lambda^2}\right]+\frac{\eta}{\mathcal{R}}\delta\left(\frac{\Upsilon}{2\eta\lambda}+\frac{\mathcal{R}\partial_{\phi}\eta}{\lambda\eta^2}\right)\Biggr\}.
\end{split}
\end{equation}
Hence the pre-symplectic form \cite{Lee:1990nz} can be written as
\begin{equation}\label{presymplectic-Y}
    \Omega [\delta g ,\delta g; g] =  \frac{1}{16 \pi G} \int_{\mathcal{C}_r} \d v \d{}\phi\,\left[ \delta(\lambda^{-1})\wedge\delta \hat{\mathcal{M}}+\delta\mathcal{U}\wedge\delta\hat{\Upsilon}+{\cal D}_v ( \delta \Omega \wedge \delta \Pi ) \right] \, 
\end{equation}
The pre-symplectic form \eqref{presymplectic-Y} still involves off-shell quantities and  should be computed over solutions of \eqref{G-N-T-metric''}.  The charge analysis shows that in $3d$ we can have at most four co-dimension one charges, namely $\{\hat{\mathcal{M}}, \hat{\Upsilon}, \Omega, \Pi \} $, in the adopted coordinate system. $\lambda$ and $\mathcal{U}$ do not lead to independent charge variations but the symplectic form assure that they do not correspond to degeneracy directions on the phase space. 

The last term in \eqref{presymplectic-Y}  involves $\Omega$ and its canonical conjugate $\Pi$ is a total $v$ derivative.  We note that this part involves charges which form the Heisenberg algebra in the algebra of charges.\footnote{The same is also true for the $2d$ case,  \eqref{symplectic-form-m}.} Therefore, this term may be absorbed into a $Y$-term. Such $Y$-terms were dubbed as `corner term' \cite{Freidel:2021fxf, Ciambelli:2021vnn, Freidel:2021cjp, Ciambelli:2021nmv, Freidel:2021dxw, Speranza:2022lxr}. 

\paragraph{Direct sum Virasoro slicing.}
For (A)dS$_3$ spacetimes for which $\Lambda =-\frac{1}{\ell^2} \neq 0$ (dS$_3$ can be achieved by analytic continuation of $\ell$ to imaginary numbers, $\ell \to i \ell$) one can consider a simple change of basis,
\begin{equation}
    \mathcal{L}_{\pm}(v,\phi):= \frac{1}{16 G} \left(\ell \hat{\mathcal{M}}(v,\pm \phi) \pm \hat{\Upsilon}(v,\pm\phi) \right)\, ,\qquad \mathcal{S}:= \frac{\Omega}{8G} \,, \qquad \mathcal{P}=\frac{\Pi}{8G}\, ,
\end{equation}
and rewrite the pre-symplectic form \eqref{presymplectic-Y} in terms of these new variables,
\begin{equation}\label{presymplectic-Y'}
    \begin{split}
    \Omega [\delta g ,\delta g; g]
        =  \frac{1}{2 \pi } \int_{\mathcal{C}_r} \d v \d{}\phi\,\left[ \delta\mathcal{U}^+\wedge\delta \mathcal{L}_+ + \delta\mathcal{U}^{-}\wedge\delta \mathcal{L}_-+ 8 G \,\partial_v \left( \delta \mathcal{S} \wedge \delta \mathcal{P} \right) \right] \, .
    \end{split}
\end{equation}
The pre-symplectic form indicates that chemical potentials conjugate to $\mathcal{L}_{\pm}$ are 
\begin{equation}
    \mathcal{U}^{\pm}(v,\phi):=  \frac{1}{\ell \lambda(v,\pm\phi)} \pm  \mathcal{U} (v,\pm\phi) \, .
\end{equation}
In the new slicing the surface charge variation is
\begin{equation}\label{surface-charge-Y-infty-Virasoro}
            {\delta} Q(\xi) :=\frac{1}{2\pi } \oint_{\cc_{r,v}} \d \phi \left(
            \hat{W}\delta\mathcal{S} +\hat{Z} \delta \mathcal{P}+\epsilon^{+} \delta\mathcal{L}_{+} +\epsilon^{-} \delta\mathcal{L}_{-} \right).
\end{equation}
where
\begin{equation}
\epsilon^{\pm}(v,\phi)= \frac{1}{\ell}\hat{T}(v,\pm\phi)\pm \hat{Y}(v,\pm\phi)
\end{equation}
Transformation laws reads
\begin{equation}
    \delta_{\xi} \mathcal{L}_{\pm}=\epsilon^{\pm}\partial_{\phi}\mathcal{L}_{\pm}+ 2\mathcal{L}_{\pm}\partial_{\phi}\epsilon^{\pm}-\frac{\ell}{8G}\partial_{\phi}^3\epsilon^{\pm}.
\end{equation}
Equations of motion yield
\begin{equation}
    \partial_{v}\mathcal{L}_{\pm}-\mathcal{U}^\pm \partial_{\phi}\mathcal{L}_{\pm}-2\partial_{\phi} \mathcal{U}^\pm\ \mathcal{L}_{\pm}+\frac{\ell}{8G}\partial_{\phi}^3\mathcal{U}^\pm =0, 
\end{equation}
and transformation laws for chemical potentials $\mathcal{U}^{\pm}$ are
\begin{equation}
    \delta_{\xi}\mathcal{U}^{\pm}=\partial_{v}\epsilon_{\pm}+\epsilon_{\pm}\partial_{\phi}\mathcal{U}^{\pm}-\mathcal{U}^{\pm}\partial_{\phi}\epsilon_{\pm}
\end{equation}
and finally the charge algebra becomes
\begin{subequations}\label{Heisenberg-direct-sum-algebra}
    \begin{align}
        &\{\mathcal{S}(v,\phi),\mathcal{P}(v,\phi')\}=\frac{\pi}{4G }\delta\left(\phi-\phi'\right)\\
        &\{\mathcal{L}_{\pm}(v,\phi),\mathcal{L}_{\pm}(v,\phi')\}= 2 \pi \left(\mathcal{L}_{\pm}(v,\phi')\partial_{\phi}-\mathcal{L}_{\pm}(v,\phi)\partial_{\phi'}+\frac{\ell}{8G}\partial_{\phi}^3\right)\delta\left(\phi-\phi'\right)\,
    \end{align}
\end{subequations}
This algebra at any constant $v$ slice is Heisenberg $\oplus$ Vir $\oplus$ Vir where Virasoros are at the Brown-Henneaux central charge \cite{Brown:1986nw}. Using different change of slicing one could construct several interesting algebra such as two copies of Heisenberg algebra or four copies of Virasoro algebra.

\subsection{Reduction to $2d$}\label{sec:reduction-2d}

In this section we discuss how the $2d$ results may be obtained upon reduction of the $3d$ over the circle parametrized by $\phi$. To this end, we start with metric \eqref{G-N-T-metric''} and  suppress the $\phi$ dependence of all metric coefficients and rename ${\cal R}$ by $\Phi(v)$. To get a $2d$ Einstein-dilaton theory we need to turn {off} the off-diagonal $v\phi$ term, i.e. $U=0$, which in turn yields ${\cal U}=0=\hat{\Upsilon}$. Upon this reduction the $3d$ theory \eqref{3D-action} reduces to $2d$ JT theory, i.e. \eqref{2D-action} with $X=2\Lambda \Phi$. In this case, \eqref{metric-components''} reduces to \eqref{Phi-Omega-2d}, \eqref{2d-solution-GF} and the $3d$ equations of motion \eqref{M-Upsilon-EoM''} yield $\partial_v \hat{\mathcal{M}}=0$ and is compatible with $\hat{\Upsilon}=0$. One can therefore replace $\hat{\mathcal{M}}=m=$constant. So, we recover the $2d$ solution space upon the reduction of the $3d$ solution space. 

One may also directly check that the $3d$ symmetry generators \eqref{null-bondary-sym-gen''} and likewise the symplectic potential \eqref{presymplectic-Y} and the $Y$-term \eqref{Y-term-3dim}  reduce to the $2d$ expression, respectively \eqref{2d-CBS}, \eqref{symplectic-form-m}, \eqref{2d-Y-term}. Therefore, the expression for the charge variations, associated changes of slicing and the charge algebras are mapped on the $2d$ expressions.

\section{Discussion and concluding remarks}\label{sec:discussion}

We constructed solution space of $2d$ and $3d$ gravity theories in presence of a causal boundary and thereby analysed symplectic form and charge variations over the solution space. We fixed the $Y$-freedom in the charge variation and the symplectic form upon the physical requirement that they should be independent of where we place the causal boundary. Explicitly, we showed there exists a covariant $Y$-term which makes the symplectic form and charge variation  $r$-independent. This feature was also observed in \cite{Geiller:2021vpg} in an asymptotic symmetry analysis in a similar $3d$ setting.  Compared to the previous null boundary analysis \cite{Adami:2020ugu}, we have {an extra} charge  associated with $r$-supertranslations,  we denoted this charge by $\Pi$. In particular, recalling  \eqref{Phi-Omega-2d} (or \eqref{calR-3d}), a shift in $r$ by $\delta r$ amounts to a shift in $\Omega$ by $\eta\lambda \delta r$. Therefore, $\Pi$ and the charge associated with superscaling in $r$ generated by the $W$ term in \eqref{2d-CBS} (or \eqref{null-bondary-sym-gen''}), as expected, form a Heisenberg algebra.  

Our analysis also uncovered the following technical points about the covariant phase space formalism and associated charge analysis.  Importantly, there are two $W$ and $Y$ freedoms in the computation of charges as well as the freedom in choosing field dependence of symmetry generators; these are not fixed by this formalism. The field dependence freedom  leads to a freedom in the definition of charges and their algebra but not the symplectic form, while the $Y$ freedom also affects the symplectic form. These freedoms may be fixed upon other physical requirements. To analyze the field dependence freedom, we have formulated changes of slicings (see appendix \ref{sec:change of slicing}). While we discussed these features in our $2d$ and $3d$ examples, we expect them to be generic to the formalism and not the specific problem analyzed here:

\begin{itemize}

\item $W$-freedom fixing.

The $W$-freedom can be used to adjust the Noether charge to become equal to the integrable part of the charges  obtained through the BT formalism. This has also been discussed by Freidel et al \cite{Freidel:2021cjp}. 

\item $Y$-freedom fixing.

As mentioned, $Y$-freedom affects the symplectic form and the charge variation while $W$-freedom does not affect the two. One may fix $Y$-freedom upon requirement that symplectic form becomes $r$ independent or to remove ``unwanted'' parts of the charge, e.g. the divergences and to regularize the charges \cite{Compere:2020lrt,Papadimitriou:2005ii,Geiller:2021vpg}. 
    \item Algebraic slicing. 
    
Adjusted Lie bracket guarantees the closure of symmetry generators algebra. Nonetheless, the results could be algebriod instead of a Lie algebra. At the level of charge analysis, even if symmetry generators form a Lie algebra, the central term may in general be field dependent and therefore, in general the BT bracket may yield algebriods rather than usual Lie algebras. However, these algebriods may turn into Lie algebras upon a change of slicing. In this work we showed that one can find such algebraic slicings in $2d$ and $3d$ gravity settings. The existence of algebraic slicing seems to be more general. It is desirable to establish this beyond specific examples.

\item Genuine and integrable slicings.

Among algebraic slicings there always exist a subset where the flux computed using BT method vanishes in the absence of genuine news ({e.g.} flux of bulk gravitons through the boundary). For the $2d, 3d$ cases {which we studied here}    there is no genuine news, the genuine slicing is hence an integrable slicing. One should however note that integrable slicings need not necessarily be algebraic slicings: Starting from a Lie algebra of integrable charges one can make change of slicing into an algebriod.


\end{itemize}

One may extend the above analysis to higher dimensions, i.e. extending the analysis of \cite{Adami:2021nnf} to causal boundaries. Moreover, given that the null boundary thermodynamics \cite{Adami:2021kvx} only relies on diffeomorphism invariance of the setting and not other details, it is plausible that there exists a thermodynamic description for generic causal boundary in $D$ dimensions. We hope to explore these directions in future publications. 

\section*{Acknowledgement}
We would like to thank Daniel Grumiller  and Celine Zwikel for  long term collaborations and discussions on related topics comments on the draft and  Mohammad  Vahidinia for discussions. MMShJ would like to acknowledge SarAmadan grant No. ISEF/M/400122. The work of VT is partially supported by IPM funds. The work of PM is supported in part by the National Natural Science Foundation of China under Grant No. 11905156
and No. 11935009. The work of HA is supported by the National Natural Science Foundation of China under Grant No. 12150410311.

\appendix

\section{Boundary charges, a quick review}\label{sec:bdryCharge}

We use covariant phase space formalism \cite{Lee:1990nz,Iyer:1994ys,Wald:1999wa}, to compute surface charges associated with the boundary symmetries of previous section. We start with the symplectic form. Consider a covariant Lagrangian together with a boundary term
\begin{equation}
    L[\varphi]= L_{0}[\varphi]+ \partial_\mu L_{\text{\tiny bdy}}^{\mu}[\varphi],
\end{equation}
where $\varphi$ denotes generic fields we have in the problem. One can read symplectic potential,
\begin{equation}\label{Theta-Y}
   \Theta^{\mu} [\varphi; \varphi ]= \Theta^{\mu}_{_{\text{\tiny{LW}}}} [\varphi; \delta\varphi]+ \delta L^\mu_{\text{\tiny bdy}}[\varphi]+ \partial_\nu Y^{\mu\nu} [\varphi; \delta \varphi],
\end{equation}
where $ \Theta^{\mu}_{_{\text{\tiny{LW}}}} [\varphi; \delta \varphi] $ is the Lee-Wald symplectic potential \cite{Lee:1990nz} and $Y^{\mu\nu}$ is a skew-symmetric tensor density of weight $+1$. This $Y$-term is not specified from the first principles of the covariant phase space formulation \cite{Iyer:1994ys} and is a freedom (ambiguity) in the analysis and one should fix it through other physical requirements, as we have done for $2d$ and $3d$ examples in the main text. The boundary Lagrangian may also be chosen freely. It is usually fixed through requirement of variational principle plus adopting certain boundary conditions. In our analysis we  primarily imposed neither and hence there remains another freedom (ambiguity) in reading the symplectic potential, the $W$-freedom.\footnote{{Here we have used the terminology of \cite{Wald:1999wa} for this kind of ambiguity/freedom. In other words, $ L_{\text{\tiny bdy}}^{\mu}=W^{\mu}$.}} {We have also fixed this freedom, upon another physical requirement.}

Using the symplectic potential one can define the symplectic form (see \cite{Iyer:1994ys} and appendix B of \cite{Adami:2021nnf})
\begin{equation}\label{symplectic-form-general}
    \Omega [\delta_1\varphi, \delta_2\varphi; \varphi]:=\int_{\cc_{r}} \d{} ^{D-1}x_{\mu}\, \omega^{\mu}[\delta_1\varphi, \delta_2\varphi; \varphi], \hspace{1 cm} \omega^{\mu}[\delta_1\varphi, \delta_2\varphi; \varphi]:=\delta_{1}\Theta^{\mu}[\delta_{2}\varphi,\varphi]-\delta_{2}\Theta^{\mu}[\delta_{1}\varphi,\varphi].
\end{equation}
Given the symplectic potential one can compute the Hamiltonian generators (charge variations) associated with the symmetry generators $\xi$ \cite{Iyer:1994ys}: 
\begin{equation}
    \slashed{\delta} Q_{\xi} [\delta\varphi, \delta_\xi\varphi; \varphi]:=\int_{\cc_{r}} \d{} ^{D-1}x_{\mu}\, \omega^{\mu}[\delta\varphi,\delta_{\xi}\varphi;\varphi].
\end{equation}
By the fact that the symplectic current is conserved on-shell, $\partial_{\mu}\omega^{\mu} \approx 0$, and by virtue of the Poincaré lemma, $\omega^{\mu}[\delta\varphi,\delta_{\xi}\varphi;\varphi]=\partial_{\nu}\mathcal{Q}^{\mu \nu}_\xi[ \delta \varphi;\varphi]$, we get
\begin{equation}\label{charge-variation}
        \slashed{\delta} Q_{\xi} = \oint_{\cc_{r,v}} \mathcal{Q}^{\mu \nu}_\xi[\delta \varphi;\varphi] \d x_{\mu \nu}
\end{equation}
where
\begin{equation}
    \mathcal{Q}^{\mu \nu}_\xi=\mathcal{Q}_{_{\text{\tiny{LW}}} \,\xi} ^{\mu \nu}+ {\cal Y}^{\mu\nu}_\xi, \hspace{1 cm} \mathcal{Y}^{\mu \nu}[\delta \varphi, \delta_\xi\varphi ; \varphi]:= \delta Y^{\mu \nu}[\delta_\xi \varphi ; \varphi]- \delta_\xi Y^{\mu \nu}[\delta \varphi ;\varphi] - Y^{\mu \nu}[\delta_{\delta\xi} \varphi; \varphi] \, ,
\end{equation}
and ${\cc_{r,v}}$ is the codimension two surface at constant $r, v$, i.e. a constant $v$ slice at the causal boundary. $\mathcal{Q}_{_{\text{\tiny{LW}}} \,\xi} ^{\mu \nu}$ can be directly read from the Lagrangian we start with. We note that the above formula for the surface charge variation which involves an integration over $\cc_{r,v}$ is written for generic dimension $D$. For the $2d$ case $\cc_{r,v}$ is a point and hence there is no integral. As such the charge variation $\slashed{\delta} Q_{\xi}$ in general is a function of $v$ and $r$. Since the charge variation involve $\delta \Theta^\mu$, the boundary term $L_{\text{\tiny bdy}}^{\mu}$ (and hence $W$-freedom) do not contribute to the charge variation computed within the covariant phase space formalism, while $Y$-term affects it. 

\paragraph{Barnich-Troessaert (BT) charge-flux splitting.} 

The charge variation \eqref{charge-variation} may not be integrable, i.e. there may not be charges $Q_\xi(\varphi)$ such that 
$\slashed{\delta} Q_{\xi}={\delta} Q_{\xi}(\varphi)$. While the (non)integrability in general depends on the adopted phase space slicing, e.g. see \cite{Adami:2020ugu, Adami:2021sko, Adami:2021nnf, Geiller:2021vpg, Adami:2020amw}, one may try to split the charge variation $\slashed{\delta} Q_{\xi}$ into an integrable part and a flux part,
\begin{equation}\label{Flux-QI}
\slashed{\delta} Q_{\xi}= \delta Q^{\text{I}}_{\xi} +{ F_\xi (\delta \varphi)}.
\end{equation}
The right-hand-side in the above has clearly an ambiguity/freedom in the charge-flux separation, one can shift $Q^{\text{I}}_{{\xi}}$ by an arbitrary function ${\cal A}_{{\xi}}$ and shift $F_{{\xi}}$ by $-\delta {\cal A_{{\xi}}}$. To fix this freedom, Barnich and Troessaert made the following proposal  \cite{Barnich:2011mi}: $Q^{\text{I}}_{\xi},\, { F_\xi (\delta \varphi)}$ should be such that they satisfy,
\begin{subequations}\label{BT-Bracket}
    \begin{align}
   &\delta_{\xi_{2}}Q^{\text{I}}_{\xi_{1}} := \left\{Q^{\text{I}}_{{\xi_{1}}},Q^{\text{I}}_{{\xi_{2}}}\right\}_{\text{\tiny BT}} {-} F_{\xi_{2}}(\delta_{\xi_{1}}\varphi) \label{2d-BT-Bracket-01}\\
     &\left\{Q^{\text{I}}_{\xi_{1}},Q^{\text{I}}_{\xi_{2}}\right\}_{\text{\tiny BT}} =\, Q^{\text{I}}_{[\xi_{1},\xi_{2}]_{{\text{adj. bracket}}}}+K_{\xi_1,\xi_2}
     \label{2d-BT-Bracket-02}
\end{align}
\end{subequations}
where $K_{\xi_1,\xi_2}$ is a possible central term and the ``adjusted bracket'' is a Lie bracket of symmetry generators $\xi$ which is adjusted for possible field dependence of the generators, see \cite{Compere:2015knw}.

\paragraph{Noether charge.} One may also compute the Noether charge. Noether method, as compared to the ones discussed above, yields the charge itself and not the charge variation. While not having the integrability issue, it is prone to $W$-freedom as well as the $Y$-freedom. From the standard Noether analysis we can read the following Noether current for diffeomorphisms
\begin{equation}
   \text{J}_{\xi}^{\mu}[\varphi]:=\Theta^{\mu}[\delta_{\xi}\varphi;\varphi]-\xi^{\mu}L[\varphi].
\end{equation}
It is easy to show that the Noether current is conserved on-shell, $\partial_{\mu} \text{J}_{\xi}^{\mu}\approx 0$,  and by using the Poincaré lemma we get, $\text{J}_{\xi}^{\mu}=\partial_{\nu}\text{N}^{\mu\nu}_{\xi}$. Considering possible contribution of $W$ and $Y$ terms, the Noether charge takes the form
\begin{equation}\label{Noether-charge}
   {\text{N}_{\xi}:= \int_{\cc_{r,v}} \d x_{\mu \nu } \, \text{N}^{\mu\nu}_{\xi} \qquad\text{with}\qquad} \text{N}^{\mu\nu}_{\xi}=\text{N}^{\mu\nu}_{\text{\tiny{LW}} \, \xi} + Y^{\mu \nu}[\delta_\xi \varphi; \varphi ] {-}2\xi^{[\mu}\mathcal{W}^{\nu]}
\end{equation}
where $\mathcal{W}^{\mu}$ is a vector density of weight $+1$. 

\section{Change of slicing}\label{sec:change of slicing}


In this appendix we review two other technical points we have used in our analysis, the notion of adjusted bracket \cite{Barnich:2011mi, Compere:2015knw} and the change of slicing over the solution space \cite{Adami:2020ugu}, see also  \cite{Grumiller:2019fmp, Adami:2021sko, Adami:2021nnf, Adami:2021kvx}. 

\paragraph{Adjusted bracket.} Let  $\xi=\xi^{\mu}\partial_{\mu}$ denote the set of diffeomorphisms  which  nontrivially act at the boundary and hence  rotate us within the solution phase space of the theory. The vector field $\xi$ (and its components $\xi^\mu$ can in general depend on dynamical fields  $\varphi$ and parameters $\mu^a$ which label the solution space, $\xi=\xi[\mu^{a};\varphi]$.  As we will review below, such field dependence naturally arise in change of slicings. Since $\xi$ move us on the solution space taking $\varphi$ to $\varphi+\delta_\xi\varphi$, when making two successive transformations one needs to account for the variation of fields in the argument of $\xi^\mu(\varphi)$ and ``adjust'' for it.  In particular, to read the algebra of symmetry generators we need to define a new, adjusted bracket \cite{Barnich:2010eb, Compere:2015knw},
\be\label{adjusted-bracket-def}
[\xi_1,\xi_2]_{_{\text{adj. bracket}}}=[\xi_1, \xi_2]-\hat\delta_{\xi_1}\xi_2+\hat\delta_{\xi_2}\xi_1
\ee
where the first term is the standard Lie bracket and $\hat\delta$ denotes variation due to field dependence of $\xi$. It is easy to show that the adjusted bracket satisfies the basic properties of a bracket. It also reduces to the usual Lie bracket when diffeomorphisms are field independent, $\hat{\delta} \xi=0$.
\paragraph{Change of slicing.} From the covariant phase space formalism, one can get the following expression for the surface charge variation associated with the symmetry generator $\xi$,
\begin{equation}
    \slashed{\delta}Q_{\xi}=\int \d{}^{D-2}x\, {\cal G}^{i} {\delta}Q_{i}\, .
\end{equation}
where ${\cal G}_i$ are  a linear combination of symmetry generators $\mu^a$ with field dependent coefficients, ${\cal G}_i=\sum_\mu  {\cal G}_{ia} \mu^a, {\cal G}_{ia}= {\cal G}_{ia}(\varphi)$.  A change of slicing,  a change of coordinate on the solution phase space, amounts to redefining $Q_i$ as
\begin{equation}
    \tilde{Q}_{i}=\tilde{Q}_{i}[Q_j, \partial^{n}Q_{j}]\, ,
\end{equation}
such that the total charge variation $ \slashed{\delta}Q_{\xi}$ remains intact. To ensure this requirement, one should transform 
${\cal G}_i$ accordingly, 
\begin{equation}
    \slashed{\delta}Q_{\xi}=\int \d{}^{D-2}x\, {\cal G}^{i} {\delta}Q_{i}=\int \d{}^{D-2}x\, \tilde{{\cal G}}^{i} {\delta}\tilde{Q}_{i}.
    \end{equation}
This requirement is fulfilled if 
\be
\tilde{{\cal G}}_{i}= {\cal M}_i{}^j\ {\cal G}_{j}, \qquad {\cal M}_i{}^j\ \frac{{\delta}\tilde{Q}_{j}}{\delta Q_k}:=\delta_i{}^k
\ee

Some comments are in order:
\begin{enumerate} 
\item Change of phase space slicing, as defined above, should not change physical observables over the solution phase space, while in general it is expected to change the algebra of charges. That is, the algebra of charges $Q_i$ and that of $\tilde{Q}_i$ are generically different. However, there could be certain changes of slicing which amount just to a change of basis of the charge algebra and does not change the Lie algebra of charges.
\item Upon a change of slicing the algebra of charges need not remain a Lie algebra. That is, commutator of charges in a new slicing may lead to a function of charges which is not a linear combination of charges. Put differently, structure `constants' of the algebra may also be field dependent. In such cases the associated symmetry generators (with the adjusted bracket) form an algebriod and not an algebra. 
\item All deformations of the algebra, see \cite{FarahmandParsa:2018ojt, Safari:2019zmc, Safari:2020pje} for discussion and analysis, are examples of changes of slicings. 
\item A change of slicing not only can change the algebra, but  it may also change the central charge.
\item More importantly, a change of slicing may take a non-integrable charge to an integrable one.
This is possible if there exists a change of slicing in which ${\tilde{\mathcal{G}}_i}$ is field independent.
In the main text we have given some examples of such integrable slicings. 
\item Existence of integrable slicings amounts to the absence of `genuine fluxes', fluxes associated with the bulk modes passing through the boundary \cite{Grumiller:2020vvv, Adami:2020ugu, Adami:2021sko, Ruzziconi:2020wrb, Geiller:2021vpg, Geiller:2021vpg, Adami:2021nnf}. One should also note that integrable slicing, if it exists, is not unique and there exists many integrable slicings.  
\end{enumerate} 

\bibliographystyle{fullsort.bst}
\bibliography{reference}

\end{document}